\documentclass[a4paper,11pt]{article}
\pdfoutput=1 % if your are submitting a pdflatex (i.e. if you have
\usepackage{jcappub} % for details on the use of the package, please
\usepackage{aas_macros} % to make journal abbreviations in AAS bib entries work

\usepackage[T1]{fontenc} % if needed
\usepackage[utf8]{inputenc}

\usepackage[english]{babel}
\usepackage{amsmath}
\usepackage{amssymb}
\usepackage{MnSymbol}
\usepackage{nicefrac}
\usepackage{amsfonts}
\usepackage{dsfont}
\usepackage{wasysym}

\usepackage{natbib}
\usepackage{hyperref}

\usepackage[markup=nocolor]{changes}

\usepackage{graphicx}% Include figure files

\usepackage{dcolumn}% Align table columns on decimal point
\usepackage{bm}% bold math

\newcommand{\bea}{\begin{eqnarray}}
\newcommand{\eea}{\end{eqnarray}}
\newcommand{\be}{\begin{equation}}
\newcommand{\ee}{\end{equation}}

\title{Quantum gravity lights up spinning black holes}
 
\author[a]{Astrid Eichhorn}
\author[b,c]{and Aaron Held}

\affiliation[a]{CP3-Origins, University of Southern Denmark,
Campusvej 55, DK-5230 Odense M, Denmark}
\affiliation[b]{Theoretisch-Physikalisches Institut, Friedrich-Schiller-Universit\"at Jena, Max-Wien-Platz 1, 07743 Jena, Germany}
\affiliation[c]{The Princeton Gravity Initiative, Jadwin Hall, Princeton University, Princeton, New Jersey 08544, U.S.}

\emailAdd{eichhorn@cp3.sdu.dk}
\emailAdd{aaron.held@uni-jena.de}

\abstract{Quantum-gravity effects in black holes are generally expected to be unobservable if they set in at transplanckian curvature scales. Here, we challenge this expectation. A near-critical spin parameter can serve as a lever arm that translates Planckian quantum-gravity effects to a global change in the spacetime: the horizon dissolves and the black hole ``lights up''.
\\
We investigate this transition between a black hole and a horizonless spacetime and find that additional lensing features appear instantaneously, when the quantum-gravity effect is added. 
In the presence of an accretion disk, a second set of internal photon rings appears in addition to the exponentially stacked set of external photon rings.
The internal and external photon rings merge into cresent-like features as a function of increasing spin parameter. We explore how these simulated images would be reconstructed by a radio-very-long-baseline-interferometry array like the Event Horizon Telescope. We find that a future next-generation Event Horizon Telescope may be sensitive to the additional lensing features. 
}

\begin{document}
\maketitle
\section{Introduction}
Does quantum gravity have consequences at large, observationally accessible, distance scales, if its typical scale is the Planck length? This question is often answered in the negative, based on the expectation that quantum-gravity effects are suppressed by (a positive power of) the ratio of the Planck length to observationally accessible length scales.
\\
Yet, exceptions to this expectation are known, in which an appropriate lever arm exists to translate Planck-scale effects into large-scale effects.
For instance, such a lever arm can be a long signal-propagation time, as is the case for symmetry-breaking effects that introduce a wavelength-dependent time lag/advance in short gamma ray bursts~\cite{Amelino-Camelia:1997ieq}, see~\cite{Addazi:2021xuf} for a review. In that case, the energy scale of the leading order effects has even been constrained to lie \emph{above} the Planck scale~\cite{2009Natur.462..331A}.
\\
A second example is known in particle physics. There, the lever arm is a \emph{logarithmic} scale-dependence of dimensionless couplings which translates $\mathcal{O}(1)$ modifications at the Planck scale into $\mathcal{O}(1)$ modifications at the electroweak scale. Quantum-gravity approaches with high predictive power, such as asymptotically safe quantum gravity, could therefore make predictions which are testable at the electroweak scale, some 15 orders of magnitude below the Planck scale, cf.~\cite{Shaposhnikov:2009pv,Harst:2011zx,
Eichhorn:2017eht, Eichhorn:2017lry, Eichhorn:2018whv,Eichhorn:2020sbo}.
\\

Within this broader context, we develop an example within black-hole physics, where the black-hole spin can serve as a lever arm to translate quantum-gravity effects at the Planck scale into large-scale effects.
Previous proposals for observational imprints of quantum gravity, such as black-hole echoes~\cite{Cardoso:2016oxy,Abedi:2016hgu,Cardoso:2016rao} and other signatures, e.g., \cite{Haggard:2016ibp,Giddings:2019jwy}, rely on horizon-scale modifications.
This would tie quantum-gravity effects to curvature scales, which for astrophysical black holes, e.g., in the LIGO mass range \cite{Abbott:2016blz} or the EHT mass range \cite{paper1,paper2, paper3, paper4, paper5, paper6, EventHorizonTelescope:2022xnr,EventHorizonTelescope:2022vjs,EventHorizonTelescope:2022wok,EventHorizonTelescope:2022exc,EventHorizonTelescope:2022urf,EventHorizonTelescope:2022xqj}, remain far sub-Planckian outside of the horizon. 

Under the assumption that quantum-gravity effects are instead tied to Planckian curvature scales, the usual expectation is that quantum-gravity effects are ``buried'' deep inside black holes and only become sizable close to classical curvature singularities.
One may thus expect that quantum-gravity effects remain inaccessible in observations of astrophysical black holes.

This expectation might break down for spinning, near-critical black holes. For these, even quantum-gravity effects at Planckian curvature scales can lead to modifications of the large-scale structure of the black hole, namely to a loss of the horizon. In turn, the absence of a horizon is observationally testable. In particular, the shadow of a black hole can disappear and the characteristic dark region in the center of a black-hole image can be replaced by an illuminated region, once the horizon is lost, i.e., the black hole ``lights up''.\\
We investigate this within the setting of regular black-hole metrics, which can be horizonless. The effect we focus on is not a ``smoking-gun'' signature of a particular quantum gravity model. Rather, it is an effect which -- if it were observed -- would be evidence that quantum gravity (or some other singularity-resolving modification of General Relativity (GR)) exists.

This paper is structured as follows: We discuss the quantum-gravity scenario in Sec.~\ref{sec:QGscenario} and corresponding modifications of the Kerr spacetime in Sec.~\ref{sec:regSpinningSpacetime}. We then determine the critical value of the spin parameter for the effect to set in and review the literature regarding theoretical and observational bounds on such high spin values in Sec.~\ref{sec:acrit}. Next, we discuss prerequisites for obtaining images in Sec.~\ref{sec:Images} and the resulting images themselves in Sec.~\ref{sec:supercritical-image-features}. We finally ask, how these images would appear if corresponding sources were observed by an EHT or ngEHT array in Sec.~\ref{sec:(ng)EHT}, before we conclude in Sec.~\ref{sec:conclusions}.

\section{Quantum-gravity scenario}\label{sec:QGscenario}
We explore the following scenario:
In regular, spinning black holes, a length scale $\ell$ is associated with the resolution of the classical curvature singularity. At finite $\ell$, the event horizon is shifted inwards; the black hole becomes more compact.\footnote{A heuristic explanation of this effect is that a resolution of the singularity requires a weakening of gravity and thus an increase in compactness, because the event horizon is a consequence of strong-field gravity \cite{Held:2019xde}. Under a simplicity assumption, which prohibits effects which would shield the weakening effects at the horizon and confine them to the inside of the black hole, singularity resolution therefore results in an increase in compactness \cite{Eichhorn:2022oma}.} At the same time, the inner (Cauchy) horizon is shifted outwards. (In the case of spherical symmetry, a Cauchy horizon appears.) As $\ell$ increases, the horizons approach each other. At a critical value of $\ell$, the two horizons merge and a horizonless spacetime is left behind. 
At low and moderate values of the spin parameter, the required value of $\ell$ is much larger than the Planck length; in turn Planck-scale effects do not lead to large-scale changes of the structure of the spacetime. This is different for highly spinning black holes. For near-critical Kerr black holes, the event horizon and the inner horizon already lie very close to each other. Thus, a Planck-scale modification can suffice to turn a highly spinning black hole into a horizonless spacetime. One might thus imagine a process in which a highly spinning regular black hole is spun up by absorbing matter with orbital angular momentum. A tiny increase can then suffice to trigger this transition, at which point the black hole becomes transparent to radiation and ``lights up''.
For black holes, the lever arm that produces large-scale consequences of Planck-scale effects is therefore spin.
\\

One may wonder whether this effect is specific to quantum gravity. After all, overspinning a classical Kerr spacetime also leads to a horizonless spacetime. However, an overspun Kerr spacetime contains a naked curvature singularity and violates the cosmic censorship conjecture
\cite{Wald:1997wa}.\footnote{We use the term cosmic censorship conjecture in reference to naked singularities and not in reference to a breakdown of the Cauchy problem.} We assume that cosmic censorship must not be violated. Thus, only regular black-hole spacetimes can be overspun. In turn, a regularization of curvature invariants is an effect that is expected of quantum gravity.

\section{Regular, spinning spacetimes from the locality principle or asymptotically safe quantum gravity}
\label{sec:regSpinningSpacetime}

We work with a member of the family of regular, spinning spacetimes put forward in \cite{Eichhorn:2021iwq}. Our considerations generalize beyond this family and hold for those regular, spinning black holes, in which two horizons\footnote{Generically, the inner horizon is expected to be unstable \cite{Carballo-Rubio:2018pmi,Carballo-Rubio:2019nel}, however see \cite{Bonanno:2020fgp} and also \cite{Carballo-Rubio:2022kad} for a recent counterexample.} approach each other, both as a function of spin and and as a function of a new-physics parameter $\ell$, see, e.g., \cite{Bambi:2013ufa,Azreg-Ainou:2014pra,Ghosh:2014pba,Ghosh:2014hea,Toshmatov:2014nya,Dymnikova:2015hka,Torres:2017gix,Simpson:2021dyo,Mazza:2021rgq}.
We work in horizon-penetrating coordinates $(u, r, \chi, \phi)$, where $\chi = \cos \theta$, with $\phi$ the polar and $\theta$ the azimuthal angle, $r$ the radial coordinate, and $u$ a light-cone coordinate. The line element reads
\bea
\label{eq:regular-spinning-BH-metric-ingoing-Kerr}
ds^2 &=&-\frac{r^2-2M(r, \chi) r +a^2 \chi^2}{r^2+a^2 \chi^2}du^2 +2\,du\, dr - 4\frac{M(r,\chi) a r}{r^2+a^2\chi^2}\left(1-\chi^2 \right) du\, d\phi\nonumber\\
&{}&- 2a\left(1-\chi^2 \right)dr\, d\phi + \frac{r^2+a^2\chi^2}{1-\chi^2}d\chi^2 \nonumber\\
&{}&+ \frac{1-\chi^2}{r^2+a^2\chi^2}\left(\left(a^2+r^2\right)^2 - a^2\left(r^2-2M(r, \chi)r+a^2 \right)\cdot \left(1-\chi^2 \right) \right)d\phi^2.
\eea
It reduces to the Kerr line element for $M(r, \chi) = M = \rm const$. The mass function $M(r, \chi)$ encodes new physics beyond GR. In \cite{Eichhorn:2021iwq}, a specific class of functions was motivated by a locality principle. This principle states that the local value of the curvature invariants in the Kerr spacetime determines the size of new-physics effects. Thus, low-curvature regions are essentially unmodified from GR, whereas high-curvature regions are strongly modified. In particular, the curvature singularity at $r \rightarrow 0,\, \chi \rightarrow 0$ can be lifted. The locality principle is implemented and singularity resolution achieved, by choosing, for instance,
\be
M(r, \chi) = 
\frac{M}{
	1+ \left[
		\ell_{\rm Planck}^4 \frac{48 M^2}{(r^2+a^2 \chi^2)^3}
		\right]^{\beta/2}
}.
\label{eq:M}
\ee
Herein, $\ell_{\rm Planck}$ is the Planck scale, which was treated as a free parameter in \cite{Eichhorn:2021iwq}. The parameter $M$ denotes the mass at asymptotic infinity.
The geometry is everywhere-regular (in the sense of finite curvature invariants) for $\beta\geqslant 1$. In the following, we will use the critical value $\beta=1$. As discussed in \cite{Eichhorn:2021iwq}, curvature invariants are non-singular but can be multivalued at $r = 0 = \chi$ for this choice. Qualitatively, $\beta =1$ and $\beta=2$ lead to similar results for the present purposes; we focus on $\beta=1$ because this can more directly be connected to asymptotically safe quantum gravity.

Asymptotically safe quantum gravity \cite{Reuter:2012id,Percacci:2017fkn,Eichhorn:2018yfc,Reuter:2019byg,Bonanno:2020bil} is expected to resolve curvature singularities \cite{Bosma:2019tbl}.
Black holes inspired by asymptotically safe quantum gravity~\cite{Bonanno:1998ye,Bonanno:2000ep,Bonanno:2006eu,Falls:2010he,Falls:2012nd,Litim:2013gga,Koch:2013owa,Koch:2014cqa,Kofinas:2015sna,Torres:2017gix,Torres:2017ygl,Pawlowski:2018swz,Adeifeoba:2018ydh,Platania:2019kyx,Borissova:2022jqj} incidentally also satisfy the locality principle, at least in the spherically symmetric case and for some of the constructions in the axisymmetric case \cite{Held:2019xde}, whereas the locality principle has been disregarded in \cite{Reuter:2010xb, Litim:2013gga}.
Thus the line element \eqref{eq:regular-spinning-BH-metric-ingoing-Kerr} can also be understood as a black hole inspired by asymptotically safe quantum gravity. In asymptotically safe quantum gravity, the Newton coupling $G_N$ is constant down to (roughly) the Planck length. Below, quantum fluctuations of gravity induce a scale-dependence in the Newton coupling, such that $G_N$ scales quadratically with a distance scale $1/k$, i.e., $G_N \sim k^{-2}$. This scaling comes about because in physical terms asymptotic safety corresponds to an enhanced symmetry, namely quantum scale symmetry. It implies that any dimensionful quantity must scale according to its canonical dimensionality. Black holes inspired by asymptotically safe quantum gravity  are constructed by substituting $G_N = \rm const$ by the scale dependent Newton coupling, either in the action, equations of motion, or at the level of classical solutions.
(Note that in Eq.~\eqref{eq:regular-spinning-BH-metric-ingoing-Kerr}, we set $G_N=1$; it actually multiplies the mass, wherever it appears.)
Next, $k$ is identified with a physical scale in the black-hole spacetime. Given that quantum effects are expected to set in, when \emph{any individual} curvature invariant exceeds the Planck scale, an identification of $k$ with local curvature scales is reasonable.
 The construction developed in \cite{Eichhorn:2021iwq} satisfies this identification, because it relies on an envelope function for the absolute values of the two independent curvature invariants in the Kerr spacetime.
\\
At a first glance, Eq.~\eqref{eq:regular-spinning-BH-metric-ingoing-Kerr} does not appear to arise from a modification of the Newton coupling. However, in classical black-hole spacetimes, the mass always appears in combination with the Newton coupling, so that changes in $G_N$ can equally well be described by changes in $M$. Physically, this duality just encodes the fact that a weakening of gravity (as required for singularity resolution) can occur if the gravitational interaction, $G_N$, is weakened, or if the source of the gravitational field, $M$, is decreased. Thus, Eq.~\eqref{eq:regular-spinning-BH-metric-ingoing-Kerr} describes a black hole inspired by asymptotically safe quantum gravity.\\

Three comments are in order here. \\
First, the procedure described above, known in the literature as RG improvement, has its limitations, see \cite{Held:2021vwd}. The resulting black-hole spacetimes should not be viewed as derived from, but rather inspired by asymptotically safe quantum gravity. As such, they constitute useful phenomenological models.\\
Second, the metric in Eq.~\eqref{eq:regular-spinning-BH-metric-ingoing-Kerr} (with Eq.~\eqref{eq:M}) can be extended to $r<0$, \cite{Torres:2017gix,Lamy:2018zvj}, where it also contains closed timelike curves. The latter property implies that this spacetime should not be understood as a complete proposal for a black hole inspired by asymptotically safe quantum gravity: including the quantum-gravity inspired effects lifts the curvature singularity, but is not sufficient to cure all pathologies of the Kerr spacetime. There is, however, a well-motivated candidate for a mechanism to resolve the additional pathologies connected to the $r<0$ region. It consists in modifications of the geodesic equation, which are expected in the quantum-gravity regime, because of the interplay of quantum gravity with matter. 
Given that quantum-gravity effects act repulsively (to resolve curvature singularities), one may expect that null and timelike geodesics are also repulsed from the innermost region. The region $r<0$ may therefore be inaccessible to geodesics.
\\
Third, the spherically symmetric counterpart of Eq.~\eqref{eq:regular-spinning-BH-metric-ingoing-Kerr}  (or similar metrics with slightly different functions in place of Eq.~\eqref{eq:M}) have been proposed 
in other approaches to quantum gravity \cite{Modesto:2010rv,Carr:2020hiz,Ashtekar:2005qt,Gambini:2008dy,Gambini:2013ooa,Haggard:2014rza,Ashtekar:2018lag,Nicolini:2019irw,Ashtekar:2020ckv}. In those cases where upgrades to spinning spacetimes were made, they typically follow the Janis-Newman algorithm~\cite{Newman:1965tw, Drake:1998gf}. This results in a mass-function that only depends on $\chi$, not on $r$. While this violates the locality principle, and results in distinct features of black-hole shadows \cite{Eichhorn:2021etc}, for the purposes of this paper, the difference is immaterial.
Thus, the mechanism we will discuss could hold within asymptotically safe quantum gravity as well as other approaches to quantum gravity.\\

For our mechanism, it is relevant that Eq.~\eqref{eq:regular-spinning-BH-metric-ingoing-Kerr} features two horizons, as long as $\ell_{\rm Planck}$ and $a$ are not too large. These two horizons approach each other for $a \rightarrow M$ and annihilate at a critical value $a_{\rm crit}< M$, such that a horizonless spacetime is left behind. We will investigate how the transition from a black hole (with horizon) to a horizonless spacetime appears if the spacetime is illuminated by an accretion disk and its shadow is observed. First steps towards images of black holes inspired by asymptotically safe quantum gravity have previously been made in \cite{Held:2019xde,Kumar:2019ohr,Contreras:2019cmf}, where the shadow boundary was determined.

\section{Critical value of the spin parameter}\label{sec:acrit}
\subsection{Critical value of $a$ for different mass functions}
We expect that the ``lighting-up'' mechanism discussed here is not specific to just a single quantum-gravity approach. Instead, it could be a mechanism that may be universal across several quantum-gravity approaches.
Some of these settings may result in a mass function\footnote{Most studies to date focus on spherical symmetry and therefore recover the spherically symmetric counterpart of the spinning black holes we investigate.} as in Eq.~\eqref{eq:M}, others result in different mass functions. This motivates us to go beyond the mass function Eq.~\eqref{eq:M} and test, whether the choice of mass function influences the relation between $\ell_{\rm Planck}$ and $a_{\rm crit}$.

To determine $a_{\rm crit}$, we test when the horizon equation no longer has real roots. The horizon equation for our case is given by 
\be
g^{rr}(r = H(\theta))+ g^{\theta \theta} \left(\frac{dH(\theta)}{d\theta} \right)^2=0,\label{eq:horizoncond}
\ee 
where we use $\chi = \cos \theta$ and where $r=H(\theta)$ is the location of the event horizon, which in general depends on $\theta$. For our purposes, it is not necessary to solve this differential equation; instead we just consider the boundary condition that Eq.~\eqref{eq:horizoncond} is supplemented with: reflection symmetry about the equatorial plane, i.e., about $\theta = \pi/2$, implies that the second term in Eq.~\eqref{eq:horizoncond} is absent at $\theta = \pi/2$, and the horizon condition reduces to 
\be
g^{rr}(r= H(\theta=\pi/2))=0.\label{eq:horconeq}
\ee
We consider two mass functions which differ in how fast quantum-gravity effects fall off away from the center of the black hole. We introduce
\be
K= \frac{48 M^2}{(r^2+a^2 \chi^2)^3},
\ee
which encodes the local curvature\footnote{More precisely, it is an envelope for absolute values of the polynomially independent curvature invariants that characterize Kerr spacetime, see \cite{Eichhorn:2021iwq}.}, and write
\begin{enumerate}
\item[(i)] a polynomially fast fall-off, encoded in $M_{\rm poly}(r) = M\frac{1}{1+\left(\ell_{\rm Planck}^4 K \right)^{\beta/2}}$, i.e., Eq.~\eqref{eq:M},
\item[(ii)] an exponentially slow fall-off, encoded in $M_{\rm exp}(r) = M \exp \left(-\left(\ell_{\rm Planck}^4 K\right)^{\gamma} \right)$,
\end{enumerate}
where $\beta>1$ and $\gamma>1/6$ ensure that the spacetime is regular.

The value of $a$ at which the horizon dissolves -- and the black hole can ``light up'' -- depends on the mass function. We now evaluate the deviation of $|a|/M$
from 1 that is needed for the Planck-scale regularization to translate into  a loss of the horizon.
To that end, we determine the location of the horizon. We solve Eq.~\eqref{eq:horconeq} to leading order in $\ell_{\rm Planck}/M$ 
and also introduce
\be
\delta a =1- |a|/M,
\ee
such that
\be
a_{\rm crit}/M = |a|/M - \delta a.
\ee
At leading order in $\ell_{\rm Planck}/M$
and $\delta a$, we obtain
\bea
\delta a &=& 
4\sqrt{3}\,\frac{\ell_{\rm Planck}^2}{M^2}\;
+\mathcal{O}\left((\ell_{\rm Planck}/M)^4\right)
\quad\quad\quad\quad\quad\, \mbox{for }
M_{\rm poly}(r)\;,
\\
\delta a &=& 
2^{2/3}3^{1/6}\,\frac{\ell_{\rm Planck}^{2/3}}{M^{2/3}}\;
+\mathcal{O}\left((\ell_{\rm Planck}/M)^{7/3}\right)
\quad\quad\quad \mbox{for }
M_{\rm exp}(r)\;.\label{eq:deltaaexp}
\eea
These two different scalings exemplify a general point about the Planck-scale-suppression that is expected for quantum gravity: Quantum-gravity effects are expected to be suppressed by the ratio of the Planck scale to a relevant scale of the system -- in the present case, the mass of the black hole. Yet, this expectation does not specify what the suppressing function is, e.g., whether it is a power that is larger or smaller than 1, or even just a logarithmic dependence. In the present case, we find power-law-suppressions. Yet, the different powers mean that, e.g., for a supermassive black hole (SMBH) of mass $M = 10^6\, M_{\odot}$, $\delta \approx 10^{-87}$ for the polynomial mass function and $\delta \approx 10^{-29}$ for the exponential one, i.e., the difference in suppression is 58 orders of magnitude. In the present case, this difference may seem somewhat academic, given that both suppressions are substantial. Nevertheless, this highlights that some caution is necessary when stating that quantum-gravity effects are Planck-scale suppressed -- the actual numerical suppression depends on the suppressing function. 

\subsection{Astrophysical constraints and limits on spin in GR and beyond}
\label{sec:superspinning-review}
So far, our study has remained theoretical, because we have not discussed whether the near-critical spin parameter that makes ``lighting up'' of the black hole possible, can actually be realized in nature.
Thus, we review the current status of the literature on the question, whether $|a|=M(1- \delta a)$ is achievable for astrophysical black holes. We are actually not interested in the question whether a Kerr black hole can be overspun within GR. First, for our mechanism to apply, the spin can remain in the regime $|a|<M$. Second, given that the spacetime we investigate is not a solution of GR, but inspired by quantum gravity, it is unclear how informative theoretical studies finding limits on $|a|$ within GR actually are for our case.

Nevertheless, we review the situation in GR and also quote results, where available, on beyond-GR settings.
Bounds on $|a|$ which lie below $|a|=M$, which is the fundamental limit for Kerr black holes (to avoid a naked singularity), arise in distinct physical scenarios both in and beyond GR, i) absorption of a point particle, ii) black-hole mergers, iii) scattering of a field and iv)  accretion of material from a disk. 
In addition, we briefly review observations that are starting to shed some light on the distribution of spin for SMBHs.

The case of absorption of a point particle is much debated in the literature, even within GR, starting from \cite{1974AnPhy..82..548W}, which concludes that black holes in GR cannot be overspun. This result was called into question: in \cite{Hubeny:1998ga}, the author showed that even for a near-extremal Reissner-Nordstrom black hole, overcharging may be possible by throwing in a charged particle. Her result applies to a setting, in which backreaction effects can be made arbitrarily small. Similarly, \cite{Jacobson:2009kt} shows that black holes starting off with  sub-critical spin can be overspun when self-force is neglected. The result changes when self-force is included \cite{Barausse:2010ka} and \cite{Sorce:2017dst} conclude that within GR, black holes can not be overspun when test particles are thrown in.

The question has also been investigated beyond GR. 
In \cite{Cardoso:2015xtj}, the authors investigate whether the absorption of a massless point particle can overspin a black hole which is close to a Kerr black hole and find a negative answer, at least for circular black-hole spacetimes. The authors also find examples in which overspinning is possible, namely for spacetimes which feature singularities at the horizon. 
Further, \cite{Li:2013sea} investigates the absorption of a massive test particle with angular momentum for Hayward and Bardeen black holes, which are both regular. The authors conclude that such regular black holes can be overspun by absorbing a particle. Overspinning becomes easier, the larger the new-physics scale. New-physics scales of the order of the Planck scale are not explicitly investigated.
In turn, \cite{Yang:2022yvq} focuses on regular black holes in which the new-physics scale is the Planck scale, in the context of Loop Quantum Gravity. The authors find that such Loop Quantum Gravity black holes (both extremal and non-extremal ones) cannot be overspun by throwing in a test particle. \\ 
Overall, this suggests that overspinning a regular black hole may not be possible by throwing in a test particle, see also \cite{Jiang:2020mws} for the study of a specific regular black hole. It is an open question whether or not the discussed result, obtained within the setting of specific regular black holes, generalizes and stands for the regular black holes we consider here.\\

For black-hole mergers, simulations indicate final spins below $|a|=M$, see, e.g., \cite{Hemberger:2013hsa,Healy:2022wdn}.
Beyond GR, the question has not been investigated, because numerical simulations of dynamics beyond GR have not been performed systematically yet. Therefore, the question whether the merger of two regular black holes may lead to an overspun regular black hole, is open. In addition, it should be kept in mind that a given regular black hole may be a solution to a class of dynamics beyond GR; thus, to conclusively answer this question, the whole class of dynamics would have to be tested. This is a formidable task which is not likely to be completed within the near future.\\

For the scattering of a scalar field, \cite{Yang:2022yvq} finds that regular black holes arising in the context of Loop Quantum Gravity can be overspun; in contrast to the situation in GR, where this is not possible \cite{Natario:2016bay}.\\

For SMBHs, accretion from a disk is likely a key mechanism for the change in spin. Within GR, accretion from a disk has first been studied in \cite{1974ApJ...191..507T}, concluding a limit of $|a|=0.998M$. Later, \cite{Gammie:2003qi} found, performing General relativistic magnetohydrodynamic simulations, that this value may be reduced if the effects of magnetic fields are included. Within GR, accretion does therefore not appear to allow overspinning. Beyond GR, this has been investigated in \cite{Li:2012ra}, where a limit of $|a|=1.3M$ was found for the accretion from a geometrically thick disk onto non-Kerr black holes. For two regular black holes, namely the spinning Bardeen and Hayward metric, \cite{Li:2013sea} concludes that accretion from a disk can lead to overspinning. 
Given the importance of accretion for SMBHs, deriving a limit on $|a|$ from accretion onto a regular black hole as we explore here is likely most relevant. However, a proper analysis requires relativistic magnetohydrodynamic simulations that account for accretion as well as the effects of magnetic fields in the disk. These are much beyond the scope of the present paper; thus we take the existing results for regular black holes as an encouraging sign that overspinning a supermassive, regular black hole through accretion may be possible. This would make our study of potential relevance for observable SMBHs.\\

Finally, measurements can inform about typical spins, but also about whether limits inferred within or beyond GR hold.
First, the measurements of black-hole spin that arise from the observation of gravitational waves through LIGO/VIRGO \cite{LIGOScientific:2018jsj,LIGOScientific:2020kqk,LIGOScientific:2021psn} pertain to a different population of black holes than the ones that we are interested in, namely SMBHs. Thus, one should be careful when assuming that these measurements can directly inform the question whether SMBHs may reach near-critical spin values.

For SMBHs, 
\cite{Vasudevan:2015qfa} collects results on spin for 25 AGNs, finding high spin $|a|>0.9M$ in 12 of the objects, based on spectroscopy of iron lines. Within error bars, a number of spin measurements are even compatible with $|a|=M$ in that sample.
A general review of observational constraints on black-hole spin is given in \cite{Reynolds:2020jwt}, where an update of the sample in \cite{Vasudevan:2015qfa} is given, adding further SMBH candidates with high spin.
In the context of our study, it is encouraging that SMBH candidates with such high spins exist. \\

If it is not possible to reach $|a|>a_{\rm crit}$ by a physical process, nature may be said to exhibit ``quantum-gravity censorship'' in black holes: the quantum-gravity regime deep inside a black hole, as well as its observational imprint -- the ``lighting up'' process we discuss in this paper -- would not be accessible in practise. In contrast to cosmic censorship, where GR is conjectured to dynamically ``protect'' itself from naked singularities, ``quantum-gravity censorship'' would shield physically viable, non-singular spacetime regions and corresponding observational signatures. If the dynamics of quantum gravity and matter indeed prohibits a dynamical resolution of the horizon, this would be an additional challenge for the phenomenology of quantum gravity.

\section{Imaging the transition between a black hole and a horizonless spacetime}\label{sec:Images}
\subsection{Accretion disk}
To image the transition between a black hole and a horizonless spacetime, we place a non-dynamic accretion disk into the spacetime.  
We assume an optically thin disk, as expected for accretion disks of SMBHs~\cite{Johnson:2015iwg}, i.e., we neglect absorptivity.
In view of the monochromatic nature of current shadow observations, we also neglect frequency-dependence in the radiative transfer equation. We later relate image intensities at 230 and 345 GHz to each other by using a spectral index of $\alpha=1.88$~\cite{EventHorizonTelescope:2021dvx}. 

We model a stationary disk in which the rotational fluid velocity $u_\mu = \bar{u}(-1,0,0,l)$ (in Boyer-Lindquist coordinates) is determined by an angular-momentum profile $l=R^{3/2}/(1+R)$ where $R = r\sqrt{1-\chi^2}$), and we fix $\bar{u}$ such that $u_\mu u^\mu\equiv-1$, cf.~\cite[Eq.~(6-8)]{Gold:2020iql}.
The  radial disk profile is defined by a density function, cf.~\cite[Eq.~(2)]{Broderick:2021ohx},
\begin{align}
\label{eq:disk-model}
	n(r,\chi) = 
	n_0\,
	r^{-\alpha}e^{-\frac{\chi^2}{2\,h^2}}\,
	\begin{cases}
	0\;, & r<0 \\
	e^{-\frac{(r-r_\text{cut})^2}{\omega^2}}\;, & 0<r<r_\text{cut} \\
	1\;, & r>r_\text{cut}
	\end{cases}\;.
\end{align}
The parameter $r_\text{cut}$ determines the inner cutoff scale, $\omega$ determines how sharp the exponential falloff is. The disk has a Gaussian profile in vertical direction, with the height set by $h/r$ and $\alpha$ determines the radial density falloff at large distance $r\gg r_\text{cut}$.
We work with two distinct parameter choices, see Tab.~\ref{tab:parameters}, that model a disk with a relatively fast radial falloff (``fast model'') and a disk with a relatively slow falloff (``slow model''). This allows us to test the impact of distinct disk configurations on our results, in particular simulated EHT and ngEHT reconstructions in Sec.~\ref{sec:(ng)EHT}. We consider the ``fast model'' a more realistic one, because the ``slow model'' results in significant emission at relatively large distance to the black hole, which differs from what is seen in actual EHT observations.

\begin{table}[!t]
\centering
\begin{tabular}{l|c|c|c|c|}
model & $r_{\rm cut}$ & $\omega$ & $\alpha$ & $h$ \\
\hline \hline
fast model & $4M$ & $1/\sqrt{12}\,M$ & 3 & 0.1 \\
\hline
slow model & $5 M$ & $\sqrt{2}\,M$ & $0.8$ & 0.1\\ \hline  
\end{tabular}
\caption{\label{tab:parameters} We list the disk parameters for the slow and the fast model. 
}
\end{table}

To account for emission from this modelled accretion disk we numerically integrate the radiative transfer equation for the intensity~$I$, i.e.,
\begin{align}
	\frac{d}{d\lambda}\left(\frac{I_\nu}{\nu^3}\right) = C\,n\left(x^\mu(\lambda)\right),
\end{align}
along the respective null geodesics $x^{\mu}(\lambda)$. Herein, $\lambda$ is the affine parameter. 
We are only interested in the relative intensities at a single frequency. As a result, the frequency $\nu$, the dimensionful constant $C$, and the number density $n_0$ cancel out when the final images are normalized.

We obtain the images by ray tracing geodesics (backwards in time) until they either (i) cross the horizon or (ii) escape to asymptotic infinity. For horizonless spacetimes, only condition (ii) is relevant.
First, we obtain the intensity with 256-pixel resolution on a screen with screen coordinates $x,y\in[-10M,10M]$. Then, we refine the image resolution in regions of strongly varying intensity. To do so, we determine the global standard deviation of the intensity of all image points. In addition we calculate the local standard deviation of the intensity average of each image point and its nearest neighbours. Whenever the local standard deviation is larger than the global one, we locally refine the image resolution by a factor of 2. We repeat this local refinement procedure two times such that regions with large intensity variations are obtained with an effective 1024-pixel resolution.

\subsection{Photon rings}

\begin{figure}[!t]
\centering
	\includegraphics[width=0.45\linewidth]{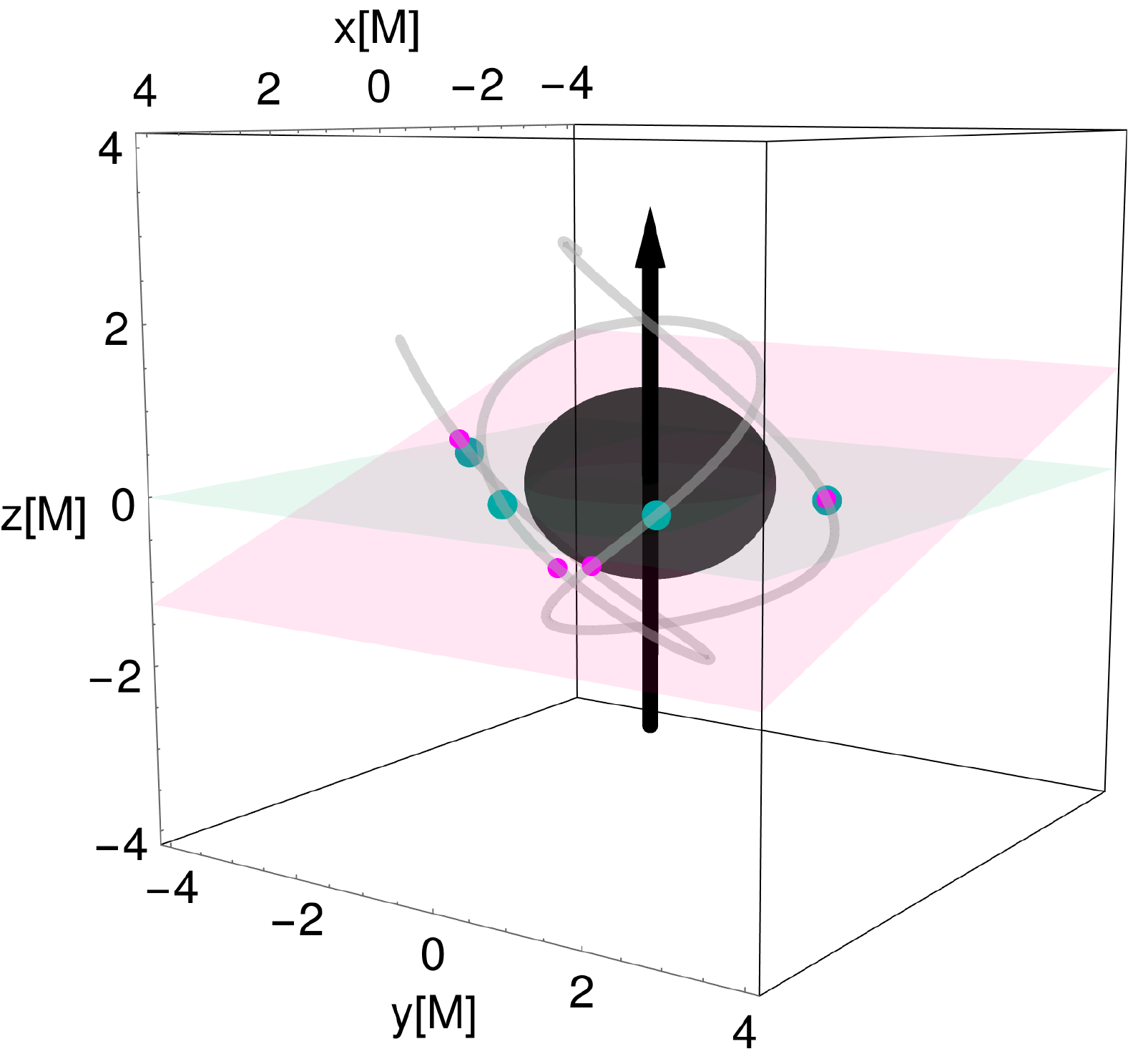}
	\hfill
	\includegraphics[width=0.45\linewidth]{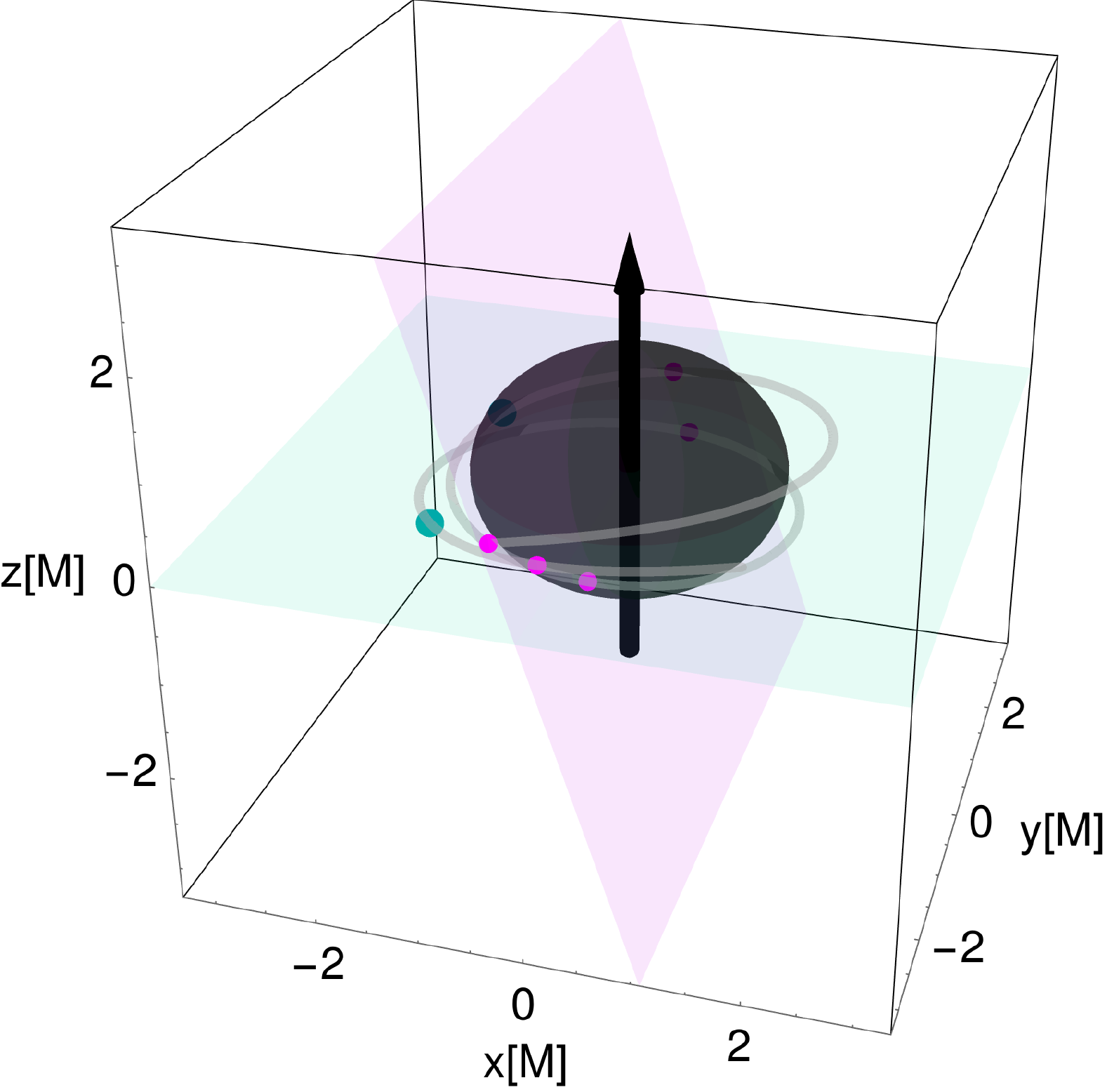}
	\\
	\includegraphics[width=0.45\linewidth]{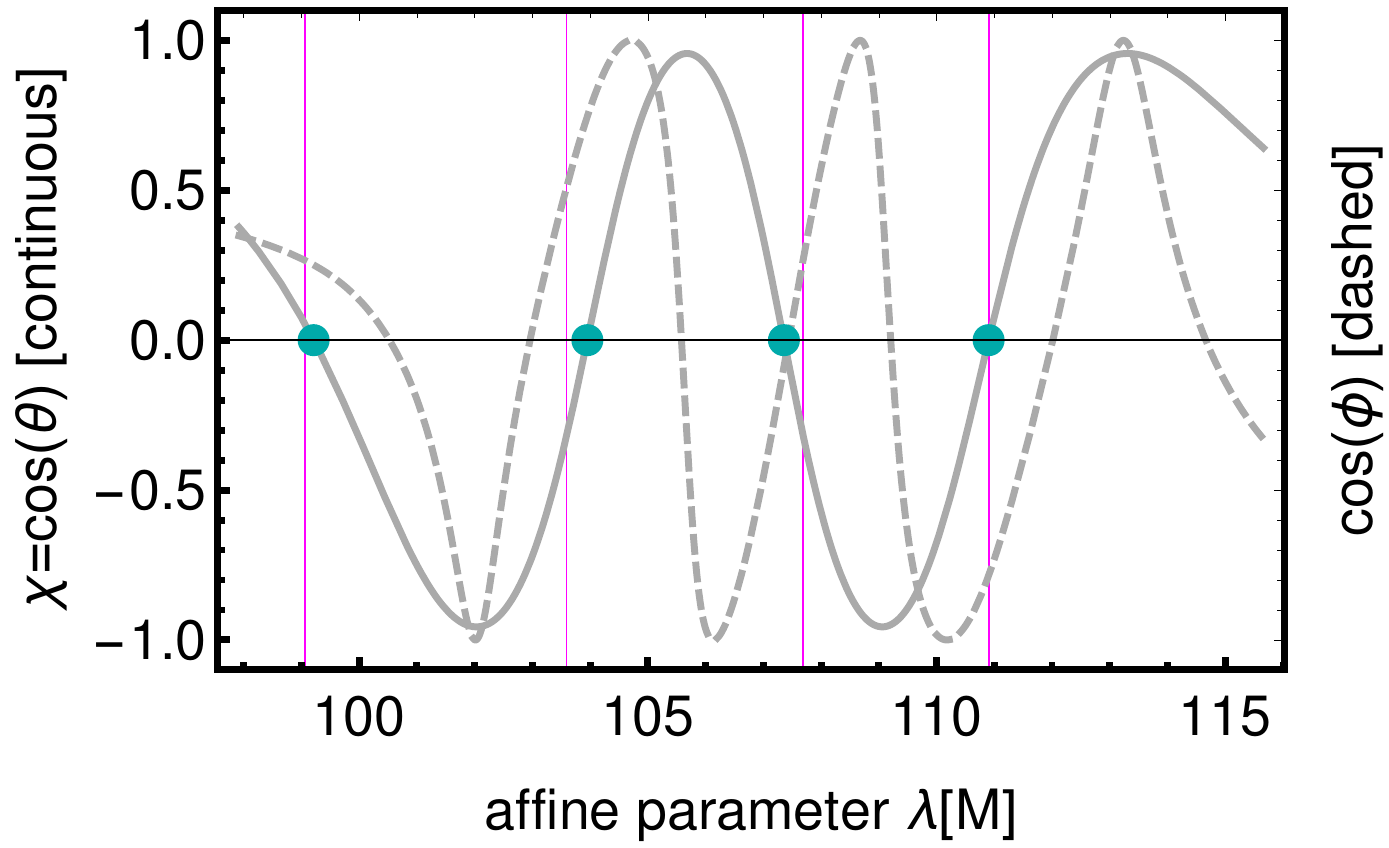}
	\hfill
	\includegraphics[width=0.45\linewidth]{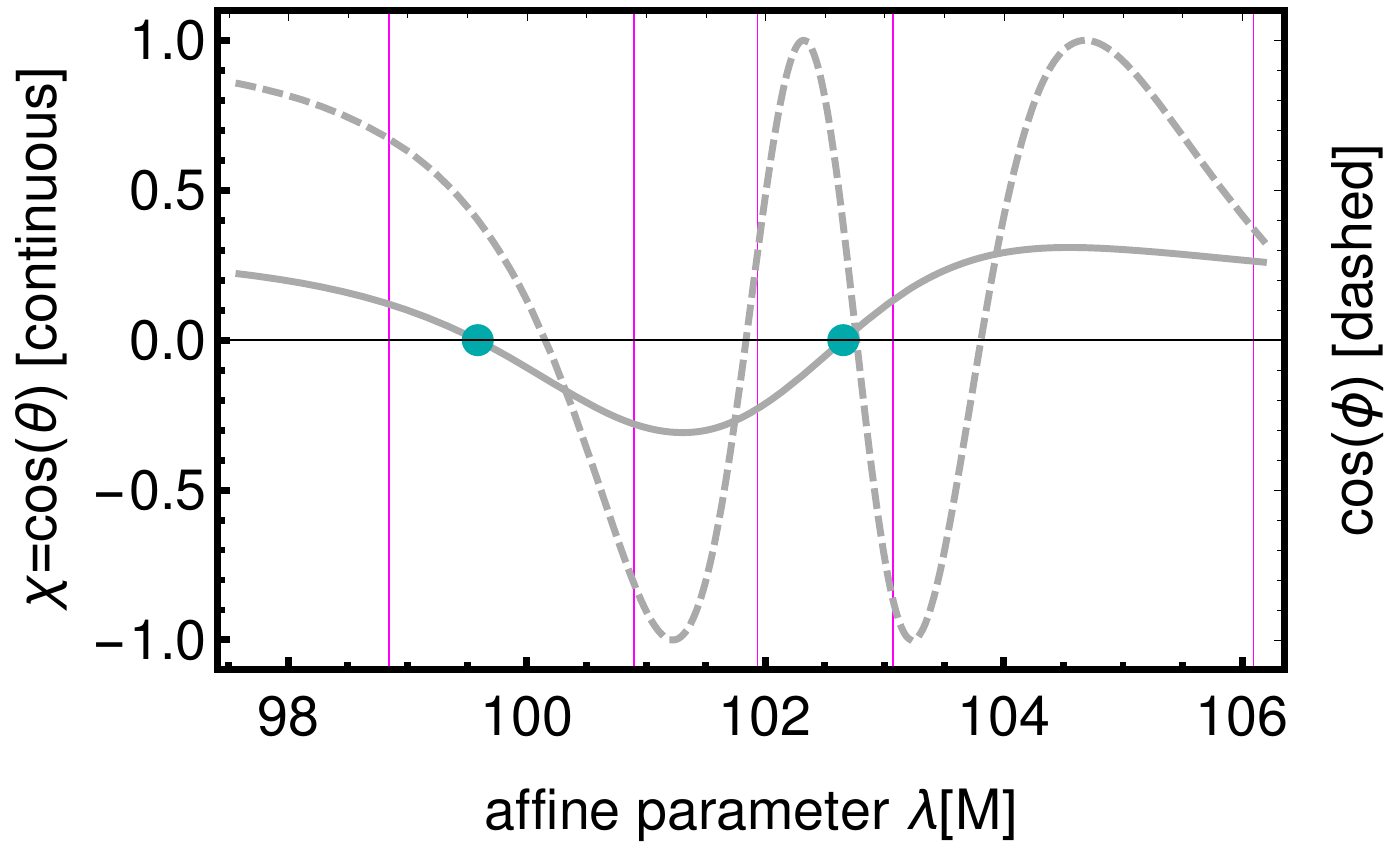}
\caption{\label{fig:ringCounting}
	We show trajectories (gray curves) for $a=0.99\,M$ in ingoing Kerr coordinates (lower panels) and in Cartesian coordinates (upper panels). 
	In the upper panels, the dark regions and the black arrows indicate the horizon and spin of the black hole; the cyan plane indicates the equatorial plane; the magenta plane indicates a plane orthogonal to the observer's line of sight; the large cyan (small magenta) dots mark where the trajectories pierce through the former (the latter) plane. 
	In the lower panels, the continuous and dashed lines correspond to (the cosine of) the polar and azimuthal angle $\theta$ and $\phi$, respectively, as a function of affine parameter $\lambda$; the cyan dots denote where $\chi\equiv\cos(\theta)=0$; the vertical magenta lines indicate values of $\lambda$ at which the trajectories pierces through the magenta plane.
	On the left (right), we show a trajectory at near face-on (edge-on) inclination, i.e., with the observer at $(r_\text{obs},\theta_\text{obs}, \phi_\text{obs})=(100M,17\pi/180,0)$ ($(r_\text{obs},\theta_\text{obs}, \phi_\text{obs})=(100M,2\pi/5,0)$), for which the piercing and orbit number agree (do not agree).
}
\end{figure}

As long as a horizon is present, the images show the expected central brightness depression, surrounded by exponentially stacked photon rings~\cite{Johnson:2019ljv}. 
These photon rings correspond to lensed images of the black hole's photon sphere. 
For the idealized case of emission arising only at infinity, they can be labelled by a winding number $n$ which counts the number of half-orbits a trajectory makes, before arriving at the imaging plane.
This generates an exponential stack of increasingly more lensed images, with $n=0$ being the emission that arises directly, $n=1$ emission that takes a ``detour behind the black hole'' and thereby completes a half-orbit and so on.

For emission arising from an equatorial disk surrounding the black hole, lensed images of the accretion disk arise, because null geodesics orbit the black hole after they have traversed the equatorial plane, where they pick up emission. In this case, careful consideration is necessary to find a definition of the winding number $n$ which labels the distinct intensity peaks in the image.\\
A priori, one could make the notion of what an ``orbit'' is, more precise in two distinct ways, cf.~Fig.~\ref{fig:ringCounting}: First, one could define an orbit number $n_o$, which is defined with the help of an observer far away from the black hole, and a plane (cutting the black hole in half) orthogonal to the line of sight of the observer to the black hole. Every time a geodesic pierces through this plane, the $n_o$ increases by 1.
However, a geodesic does not automatically traverse the disk each time it completes a full orbit that increases $n_o$ by two, cf.~right-hand panel in Fig.~\ref{fig:ringCounting}. Therefore, $n_o$ is not the relevant quantity to characterize the distinct intensity peaks in the image. Instead, an orbit is defined, see also \cite{Johnson:2019ljv}, by oscillations in $\chi$: In the Kerr spacetime, geodesics oscillate in the polar coordinate $\chi$ from some $\chi_{\rm min}$ to some $\chi_{\rm max} = - |\chi_{\rm min}|$ and back to $\chi_{\rm min}$. Thus, the disk is pierced twice (once from above and once from below) along such an orbit. Therefore, for the Kerr spacetime, counting the number of oscillations $n_{\rm osc}$ between $\chi_{\rm max/min}$ and $\chi_{\rm min/max}$ is a suitable way to count intensity peaks.

Beyond Kerr spacetime, e.g., in non-circular spacetimes, the oscillation in $\chi$ can be asymmetric, i.e., $\chi_{\rm min} \neq - \chi_{\rm max}$, see, e.g., Fig.~[1] in \cite{Eichhorn:2021etc}. Thus, it is not guaranteed that a geodesics pierces through the disk, while it completes a full orbit in $\chi$. Therefore, the definition of an orbit cannot use full oscillations in $\chi$. Instead, we define $n$ by the piercing number $n_p$: each time a geodesic traverses $\chi=0$, where it pierces through the disk, $n_{p}$ increases by one. We then relate $n$ to $n_p$ by $n= n_p-1$, such that the direct emission from the disk is associated to $n=0$.

For inclinations close to $\theta_{\rm obs}= 0$ (or $\theta_{\rm obs}=\pi$), the distinction between $n_o$, $n_p$ and $n_{\rm osc}$ disappears, as each oscillation in $\theta$ corresponds to the geodesic traversing once behind the black hole, such that an orbit has a more intuitive definition, see left-hand panel in Fig.~\ref{fig:ringCounting} and also \cite{Broderick:2021ohx}.
\\
We implement the definition of $n_p$ in our ray tracer and indeed observe that the distinct intensity peaks in the image are uniquely associated to the piercing number $n_p$, cf.~Fig.~\ref{fig:ringCounting} as well as Fig.~\ref{fig:crosssections}. In Fig.~\ref{fig:ringCounting}, we also highlight that the orbit number $n_o$ can be significantly larger than $n_p$. In the present work, where the new-physics effects are confined to the Planck scale, the piercing number $n_p$ and the oscillation number $n_\text{osc}$ are (in practise) degenerate. This changes if the new-physics effects set in at lower curvature scales, see \cite{Held:2019xde,Eichhorn:2021iwq}.

In the absence of a horizon and in the presence of a photonsphere, two 
stacked series of images make up the full image, because lensed geodesics can now also traverse the region that was previously behind the horizon, see~\cite{Lamy:2018zvj,Rahaman:2021kge,Eichhorn:2022oma,Guerrero:2022msp} for the spherically-symmetric case. Thus, geodesics that deviate slightly from the photonsphere towards large radii make up the original set of photon rings. Geodesics that deviate slightly from the photonsphere towards smaller radii make up the new set of photon rings in the inner region of the image. Thus, each ``external'' photon ring labelled by $n_\text{ext}$ comes with a twin ``internal'' photon ring labelled by $n_\text{int}$ with the same  piercing number, i.e., $n_p\equiv n_\text{ext}\equiv n_\text{int}$.
In the absence of a photonsphere, only a finite set of inner and outer photon rings exist, see \cite{Eichhorn:2022oma} and Sec.~\ref{sec:supercritical-image-features}.

\subsection{Considerations for supercritical spin and image features}
\label{sec:analyticextension}

To produce images of our spacetime, we have to make assumptions about matter densities and associated emissivities in the core region as well as the second asymptotic region at $r<0$.

In the core region, we assume vanishing emissivity and do not include any matter density.
Given that our spacetime is illuminated by an accretion disk, it is a reasonable assumption that accretion has been proceeding over a longer time scale and thus accreted matter should be present in the core region. Because we do not model the interplay of quantum gravity with matter, we do not model the state of matter in the core region. Going beyond our model of the spacetime, a time-dependent accretion process would induce a time-dependence in the metric. In \cite{Carballo-Rubio:2019nel}, it has been argued that going beyond stationarity is key to resolve pathologies (e.g., Cauchy horizons) in regular black holes. In addition, the accreted matter that may accumulate in the core region would back-react on the spacetime and change the structure close to $r=0$, such that the analytic extension may not be possible. We do not explore any of these effects, because they are not relevant for the most important image features, which arise from geodesics in the vicinity of the photonsphere and not at smaller $r$. Moreover, if a compact object formed from accreted matter radiates, that radiation is more and more redshifted, the more compact the object becomes. In addition, strong lensing effects as well as absorption of radiation by the compact object further diminish the luminosity that may be detectable \cite{Broderick:2005xa,Lu:2017vdx,Cardoso:2019rvt,Carballo-Rubio:2018jzw}, even outside the EHT band.

We do also not include any emission from the region at $r<0$.
As argued in Sec.~\ref{sec:regSpinningSpacetime}, the region $r<0$ may not be accessed by geodesics, if the geodesic equation is modified through quantum-gravity effects in the high-curvature region. We do not model such modifications here and instead work with the classical geodesic equation. Thus, geodesics can enter the region $r<0$ in our numerical treatment. We will assume, similar to \cite{Lamy:2018zvj}, that the second asymptotic region is a vacuum region and therefore does not add to the intensity of the image.

We therefore proceed as follows: In  the simulated high-resolution images (see Sec.~\ref{sec:supercritical-image-features}), we mark the region in which geodesics can enter or come from the region $r<0$. When we reconstruct VLBI observations of these simulated high-resolution images, we do not include sources of emission in the region $r<0$, such that geodesics that come from $r<0$ only pick up the emission from the accretion disk at $r>0$. The resulting image has no or negligible intensity in the image region where geodesics come from $r<0$. We consider this a good approximation of the more realistic case, where the $r<0$ region would be made obsolete due to i)  the presence of a matter core, which itself would not contribute to the detectable radiation due to redshift and/or ii) quantum gravitational effects on null geodesics, which repulse these from the Planckian core.

\section{Simulated images and their properties at supercritical spin}
\label{sec:supercritical-image-features}
\begin{figure}[!t]
\centering
	\includegraphics[width=0.325\linewidth]{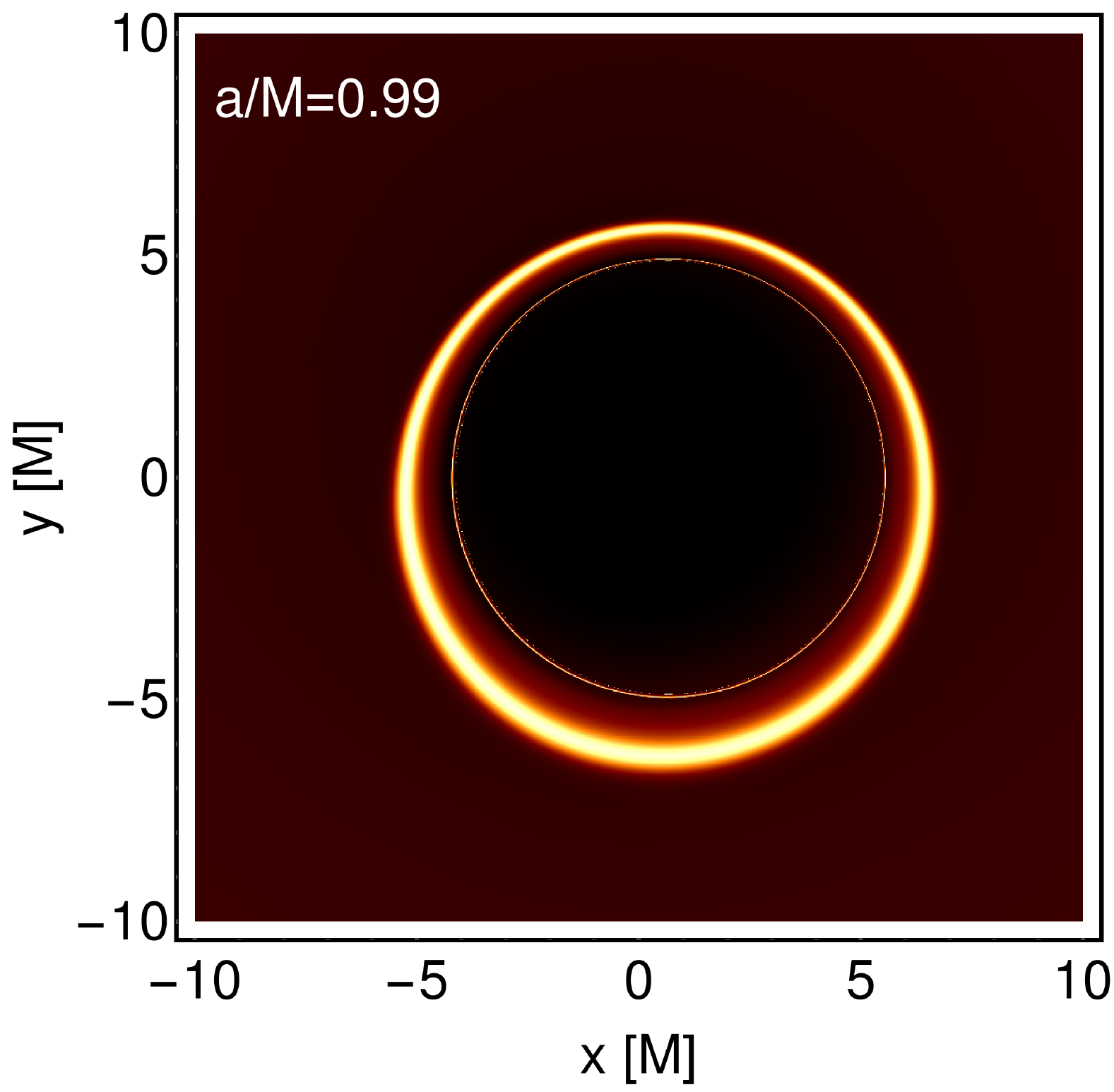}
	\hfill
	\includegraphics[width=0.325\linewidth]{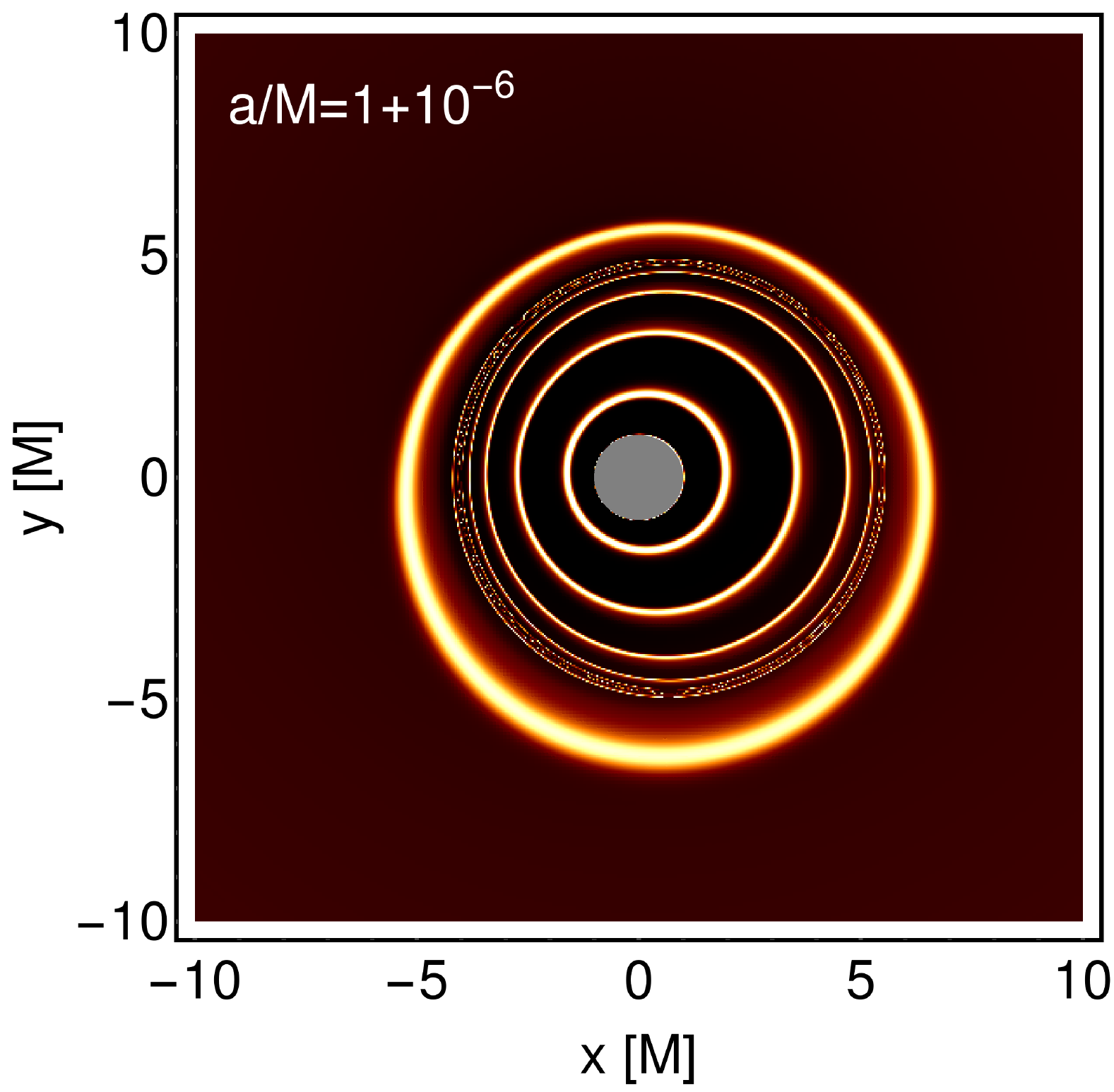}	
	\hfill
	\includegraphics[width=0.325\linewidth]{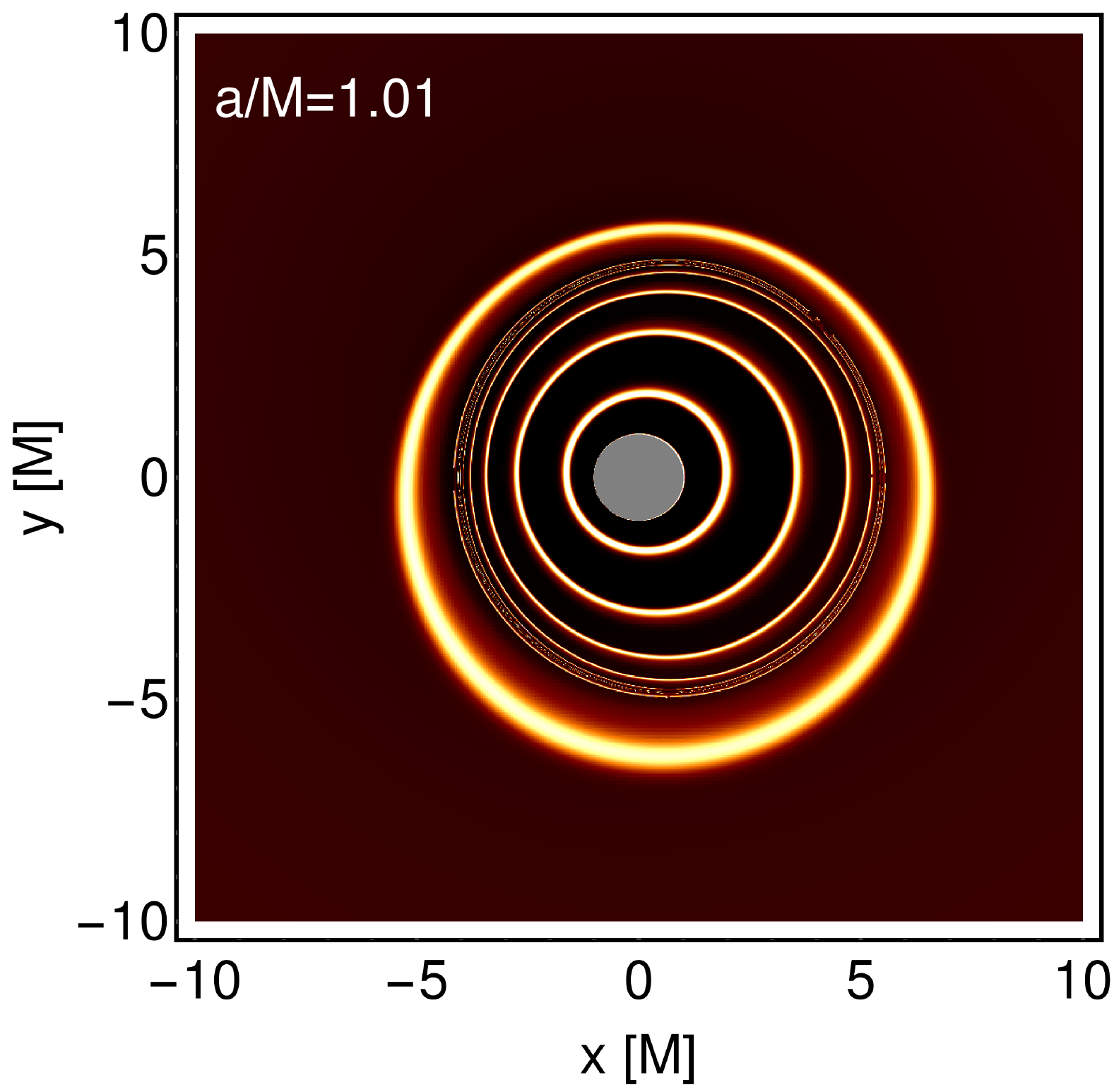}
	\\
	\includegraphics[width=0.325\linewidth]{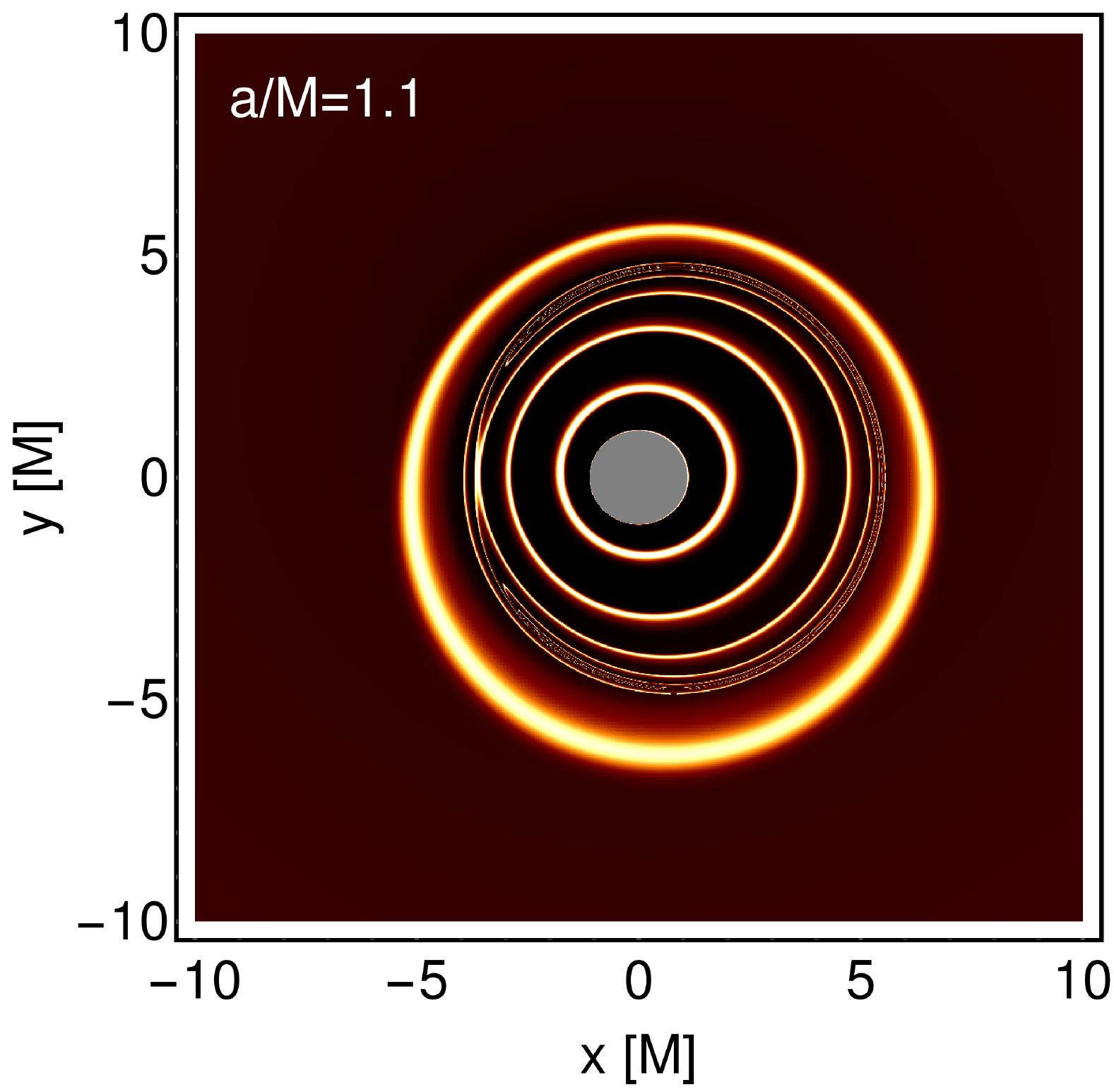}
	\hfill
	\includegraphics[width=0.325\linewidth]{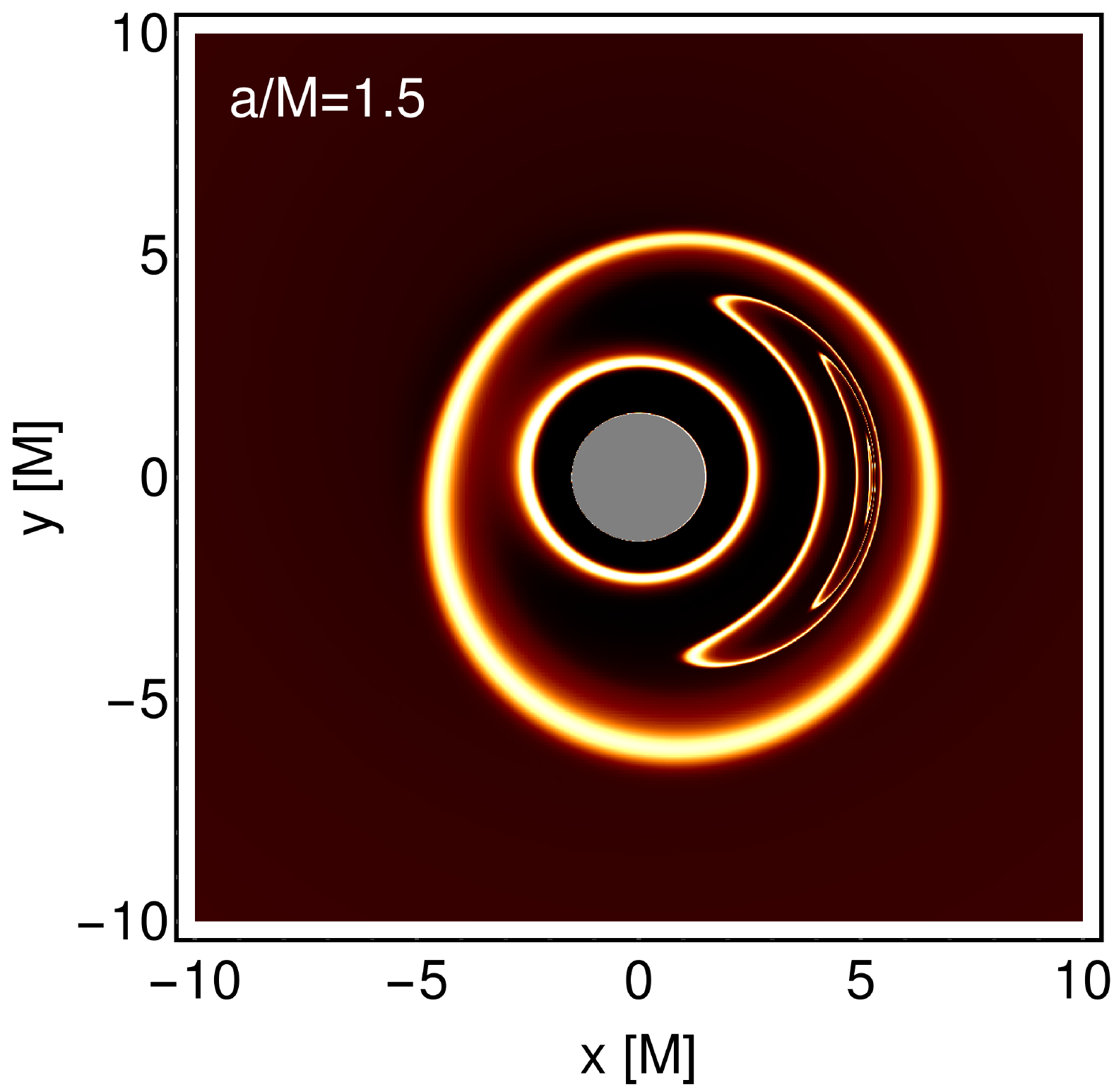}
	\hfill
	\includegraphics[width=0.325\linewidth]{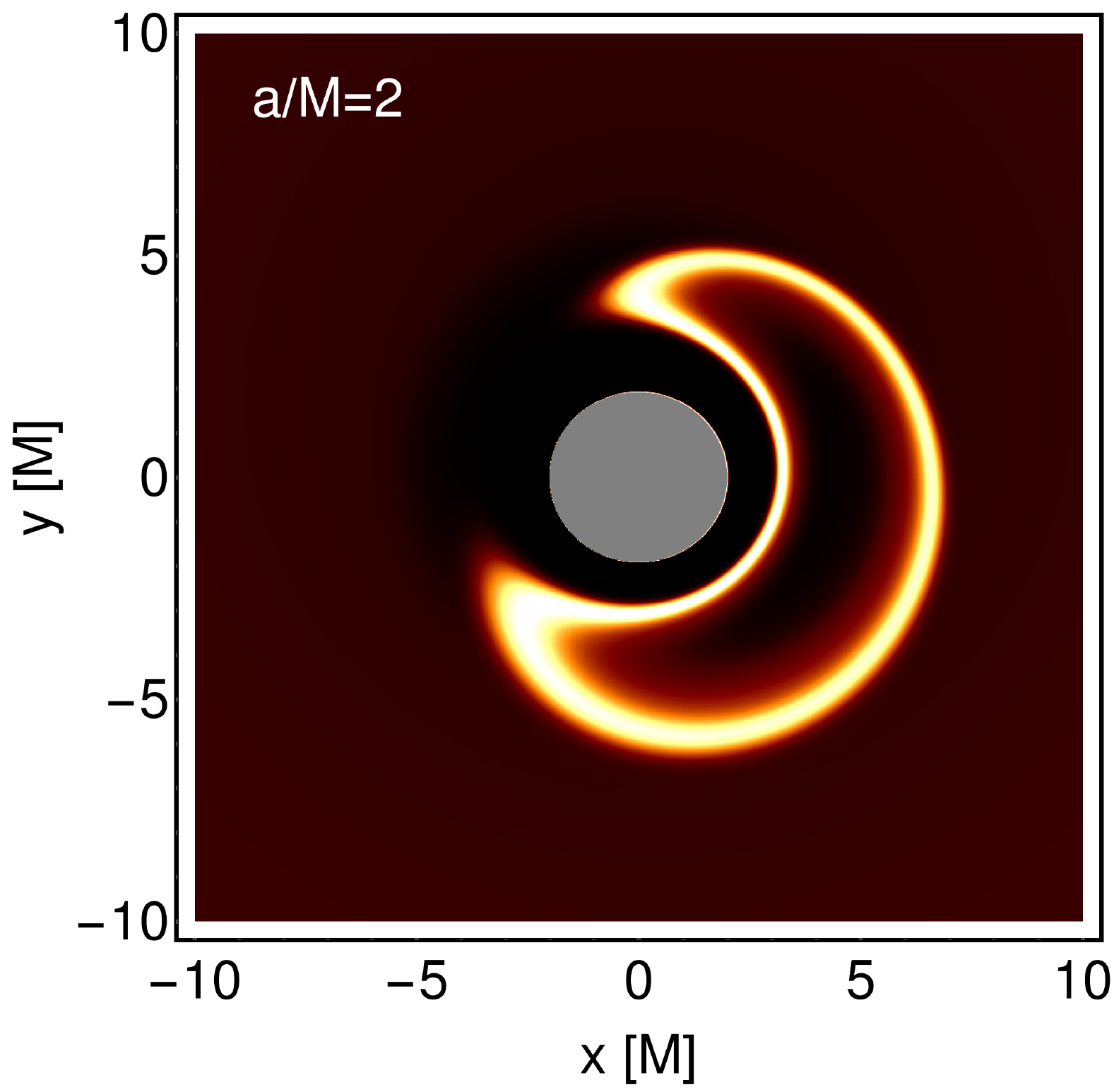}
\caption{\label{fig:superextremalM87series}We show a series of images for a regular spacetime with spin-parameter $a$, cf.~Eq.~\eqref{eq:regular-spinning-BH-metric-ingoing-Kerr}, and disk parameters as in the ``slow model'', cf.~Tab.~\ref{tab:parameters}, in the x-y image plane of an asymptotic observer at inclination $\theta_\text{obs}=17\pi/180$. From top-left to bottom-right, $a/M=0.99$, $a/M=1+10^{-6}$, $a/M=1.01$, $a/M=1.1$, $a/M=1.5$, $a/M=2$, with all but the first image being superextremal and thus horizonless. The intensity in all images is normalized to the average intensity of the $a/M=0.99$ image.
The central gray region indicates the image region in which trajectories probe the spacetime region at $r<0$.
}
\end{figure}
\begin{figure}[!t]
\centering
	\includegraphics[width=0.325\linewidth]{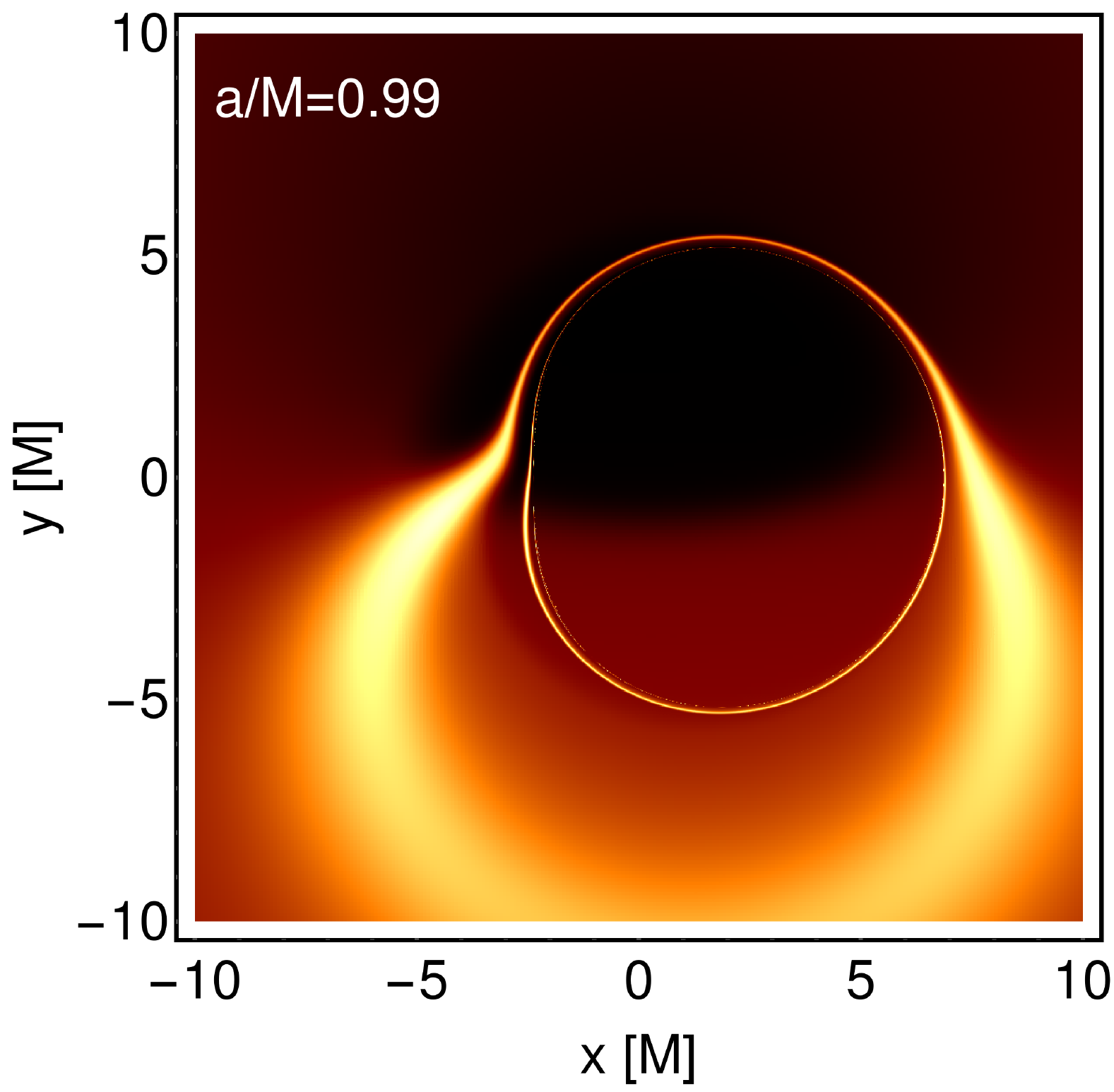}
	\hfill
	\includegraphics[width=0.325\linewidth]{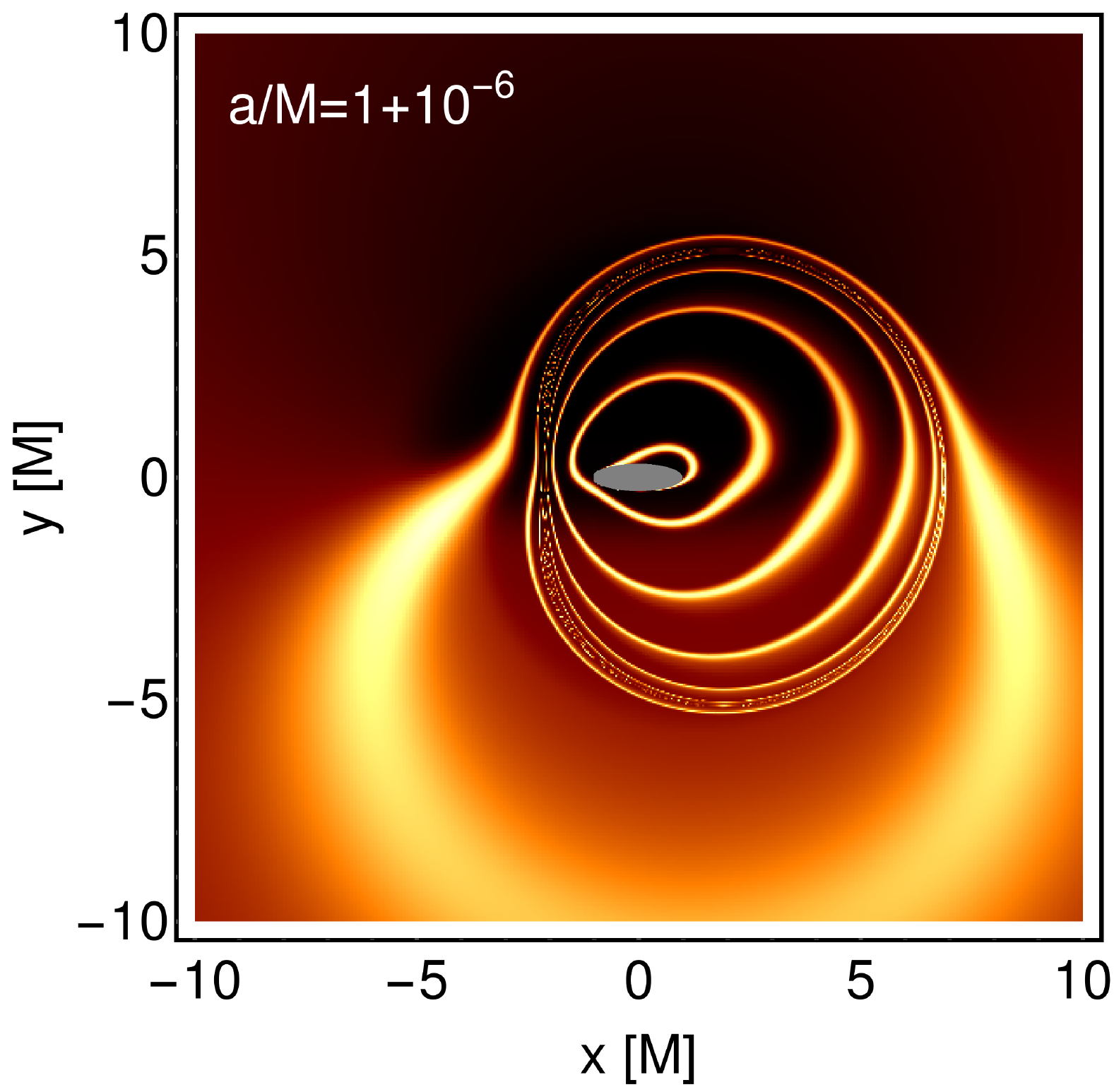}
	\hfill
	\includegraphics[width=0.325\linewidth]{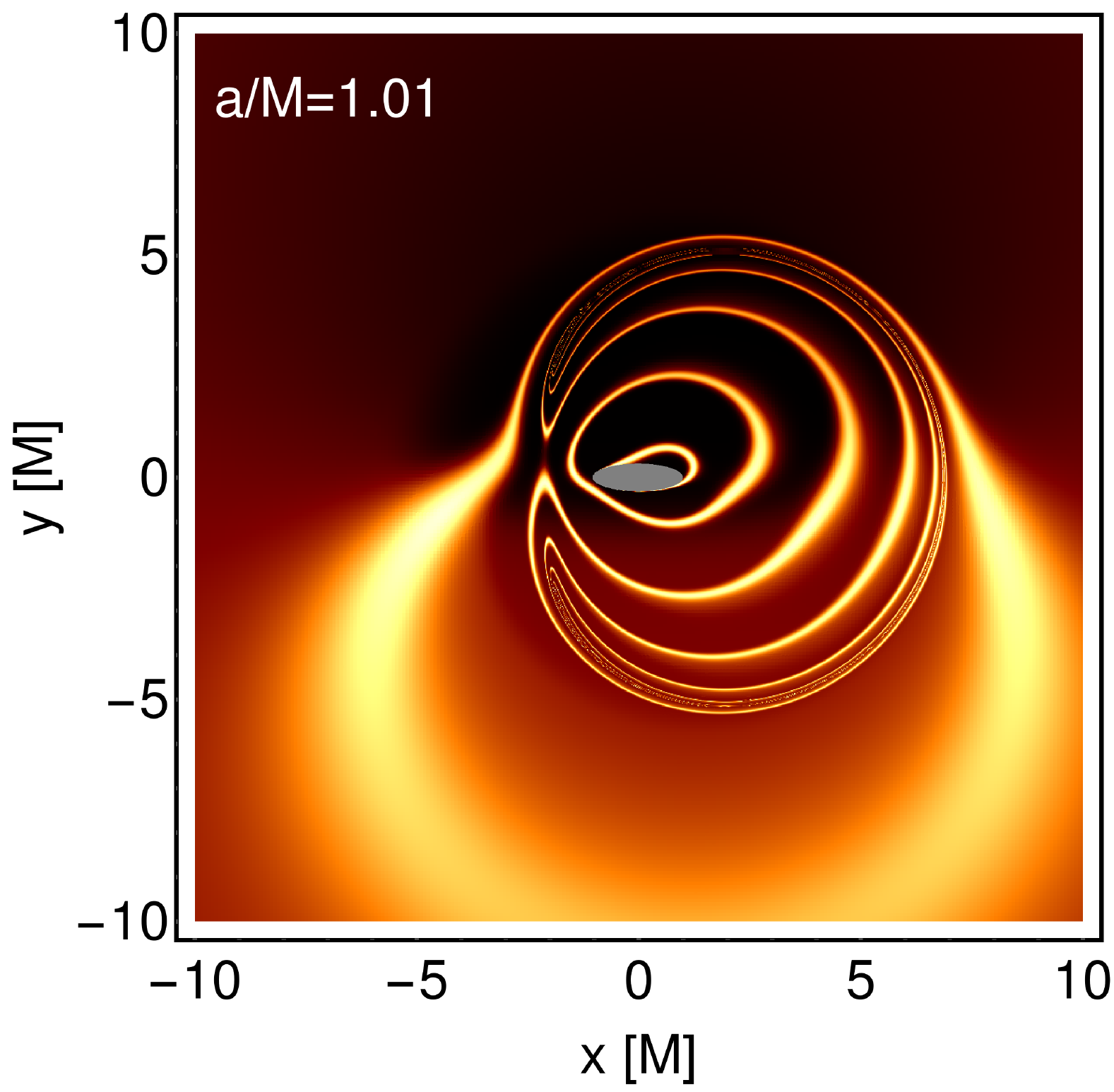}
	\\
	\includegraphics[width=0.325\linewidth]{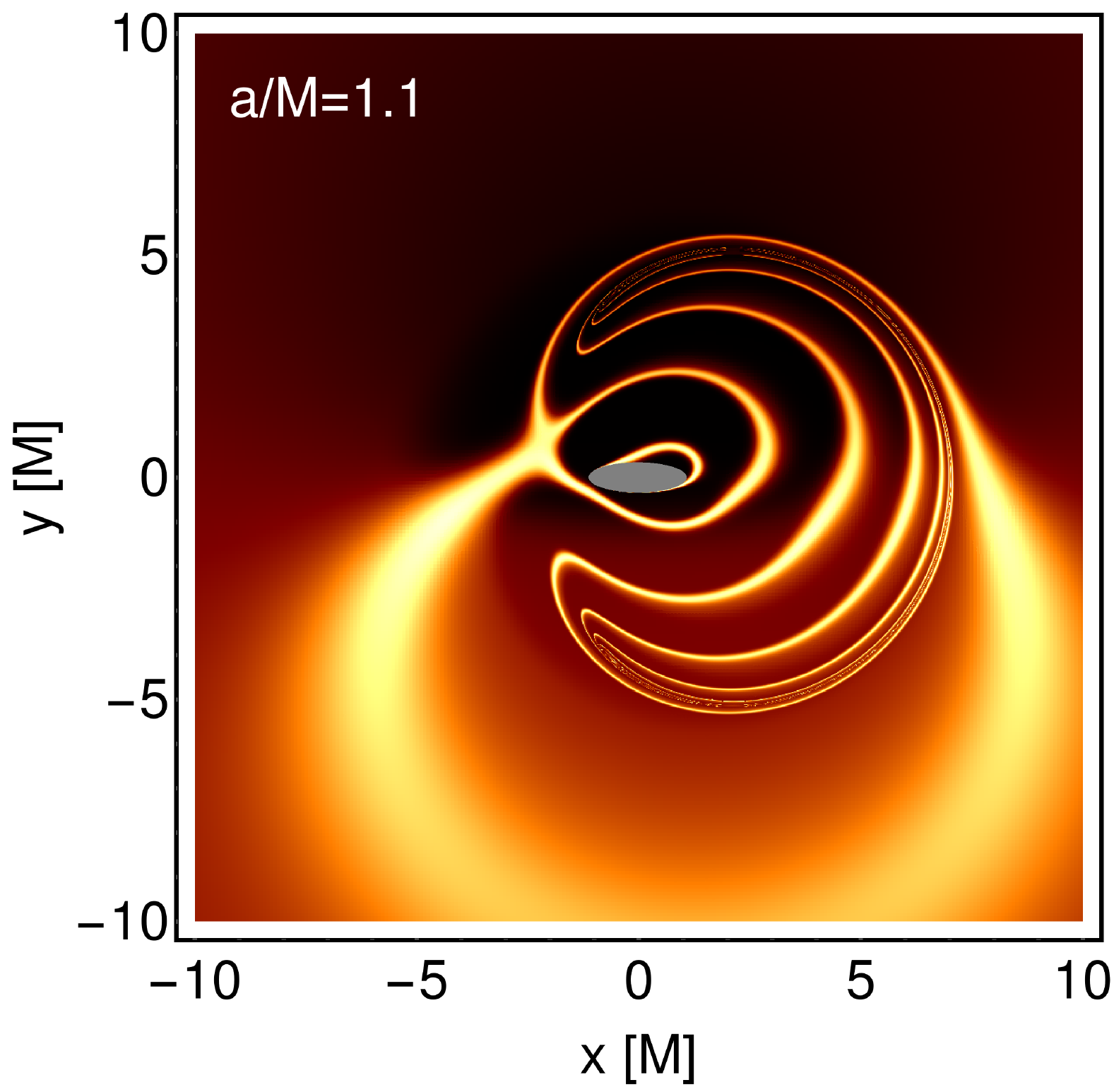}
	\hfill
	\includegraphics[width=0.325\linewidth]{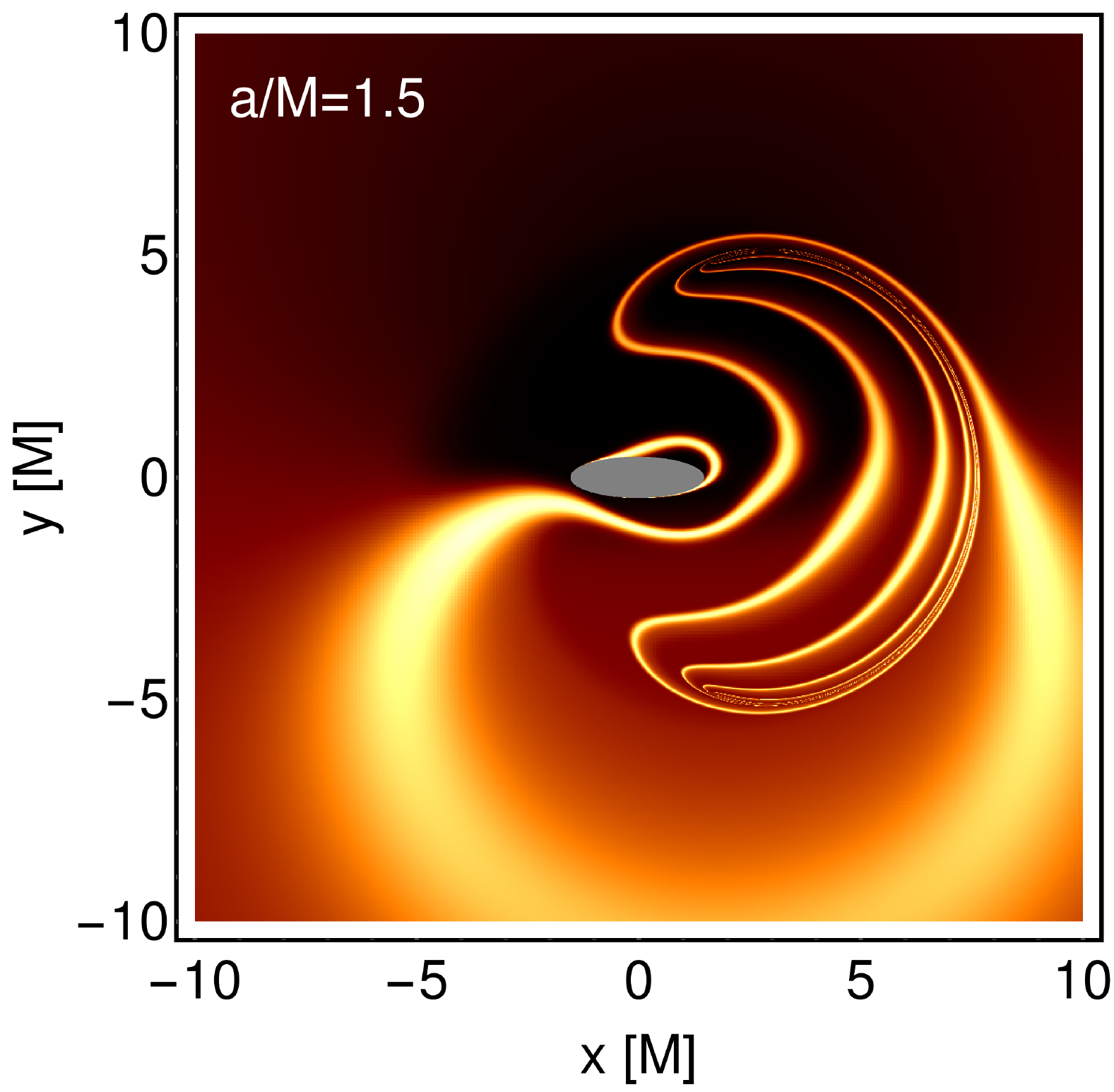}
	\hfill
	\includegraphics[width=0.325\linewidth]{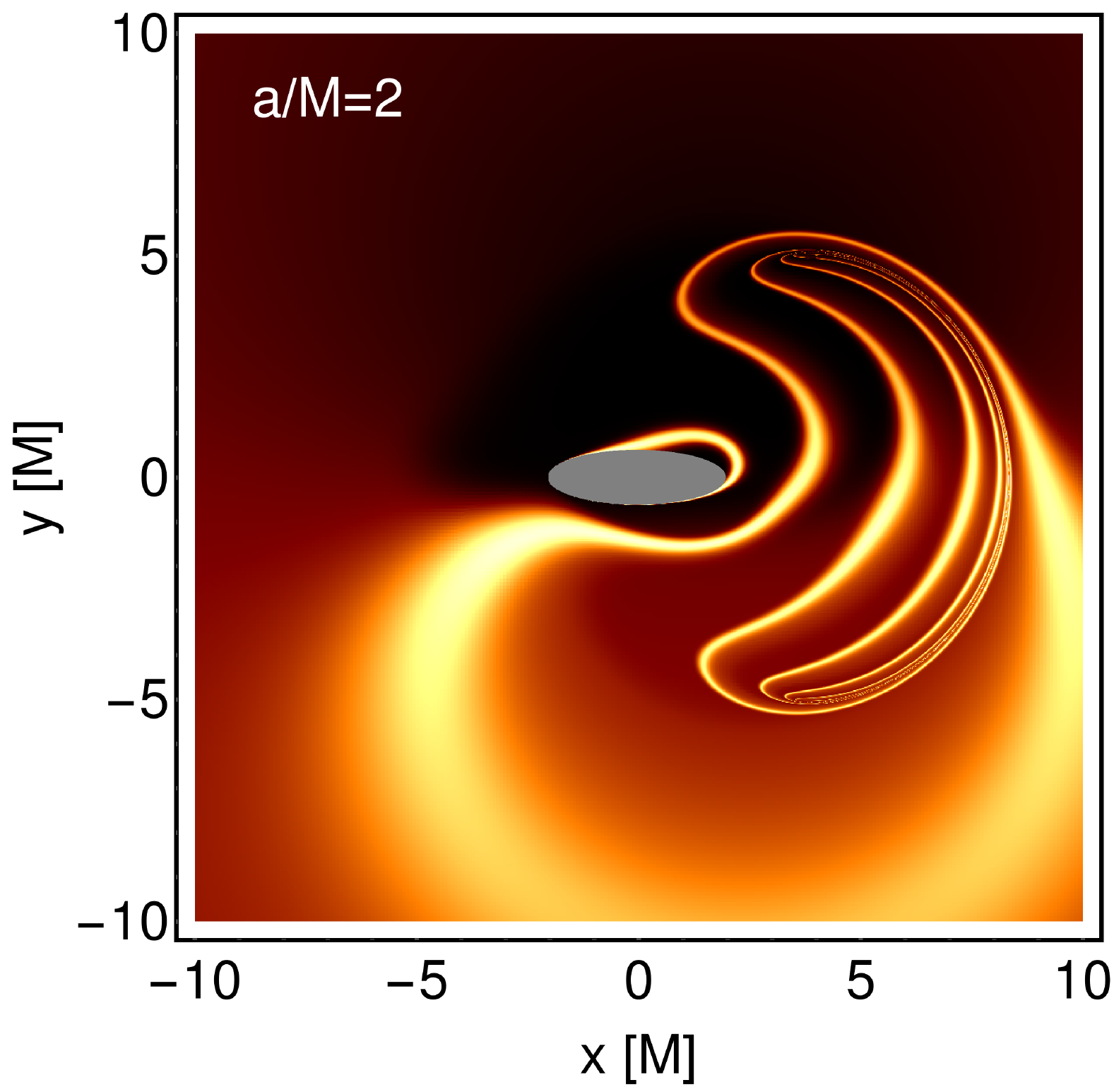}
\caption{\label{fig:superextremal8Pi20series}
As in Fig.~\ref{fig:superextremalM87series} but for near-equatorial observer inclination $\theta_\text{obs}=2\pi/5$.
}
\end{figure}
\begin{figure}[!t]
\centering
	\includegraphics[width=0.325\linewidth]{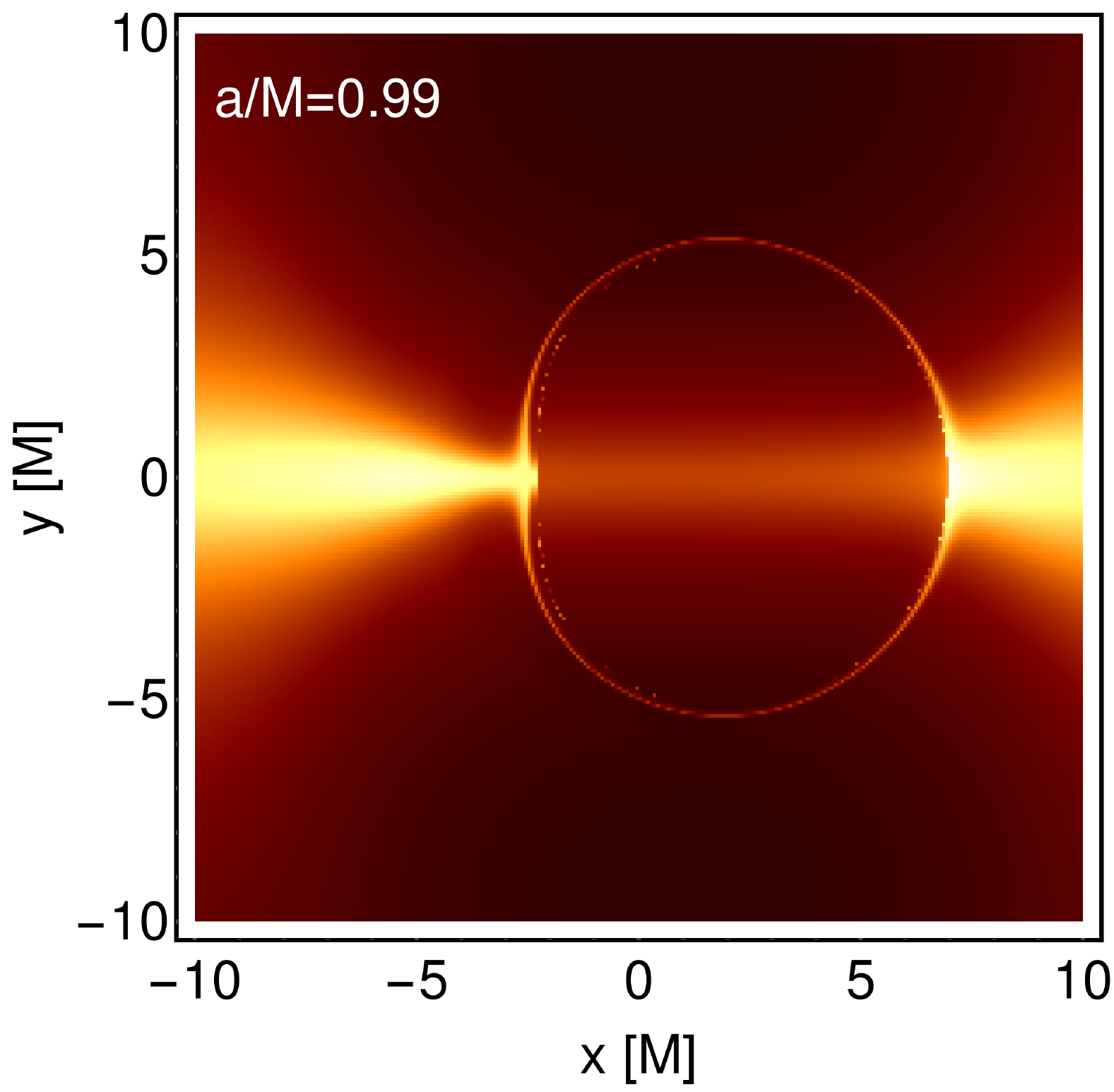}
	\hfill
	\includegraphics[width=0.325\linewidth]{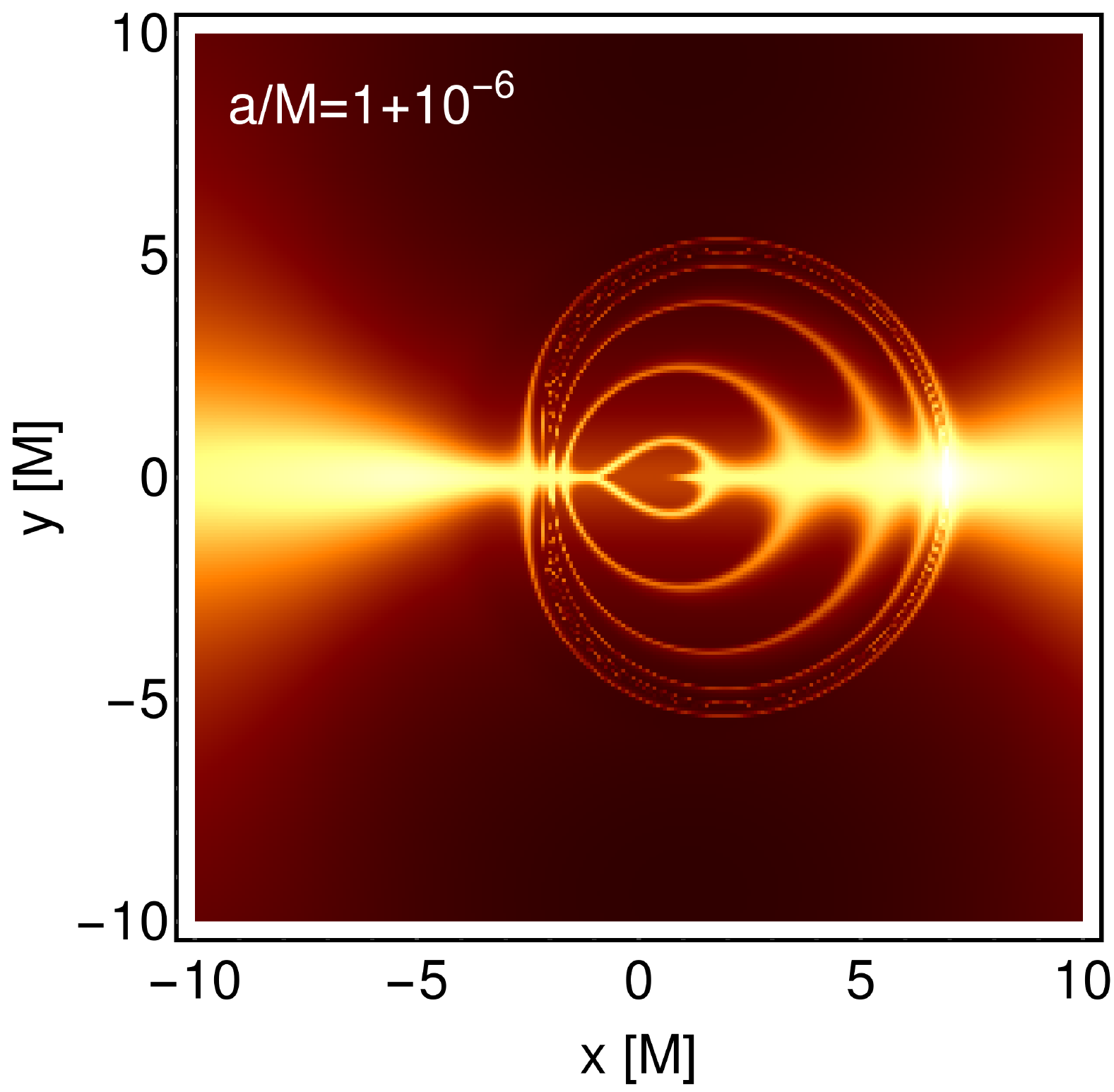}
	\hfill
	\includegraphics[width=0.325\linewidth]{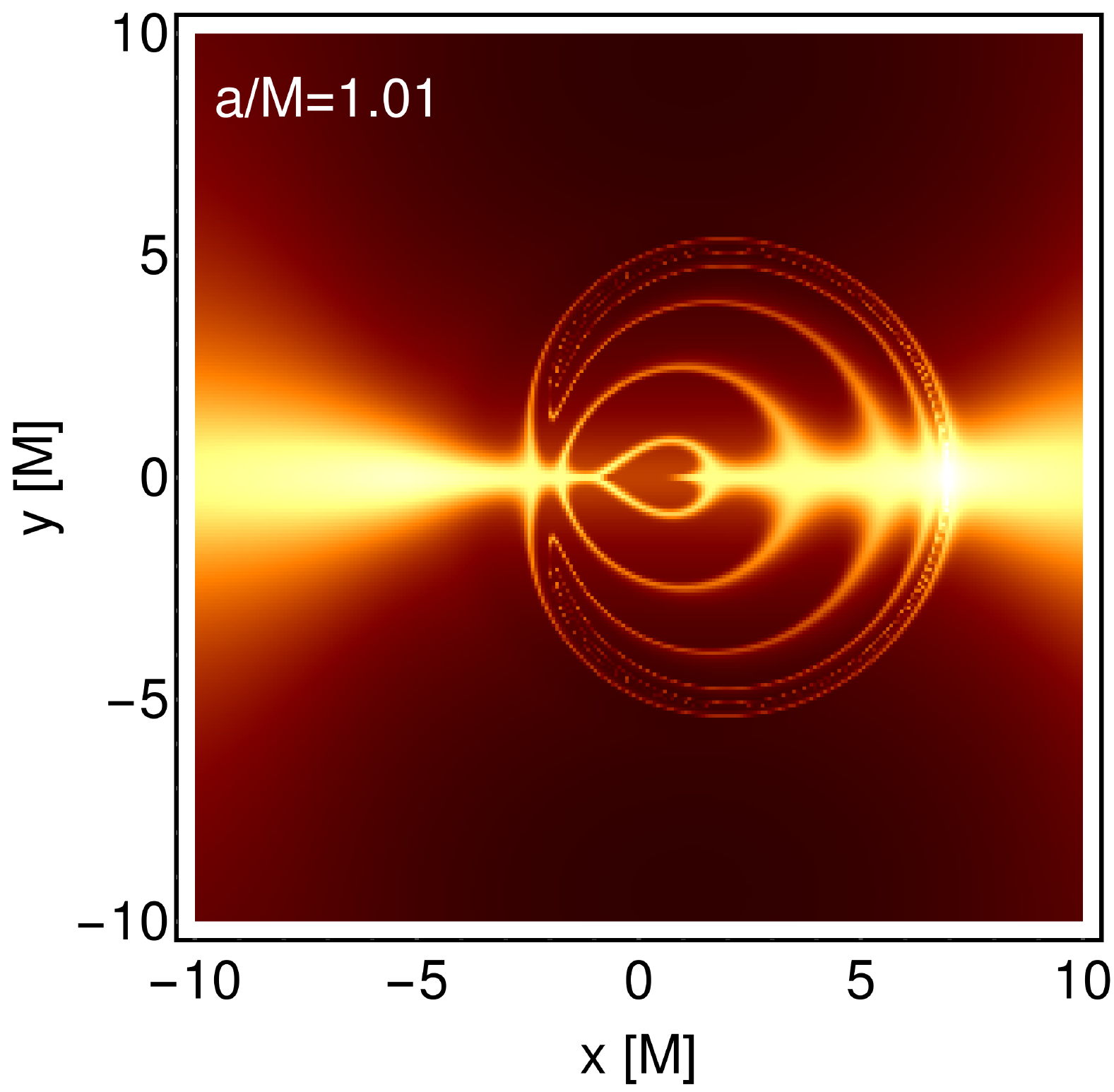}
	\\
	\includegraphics[width=0.325\linewidth]{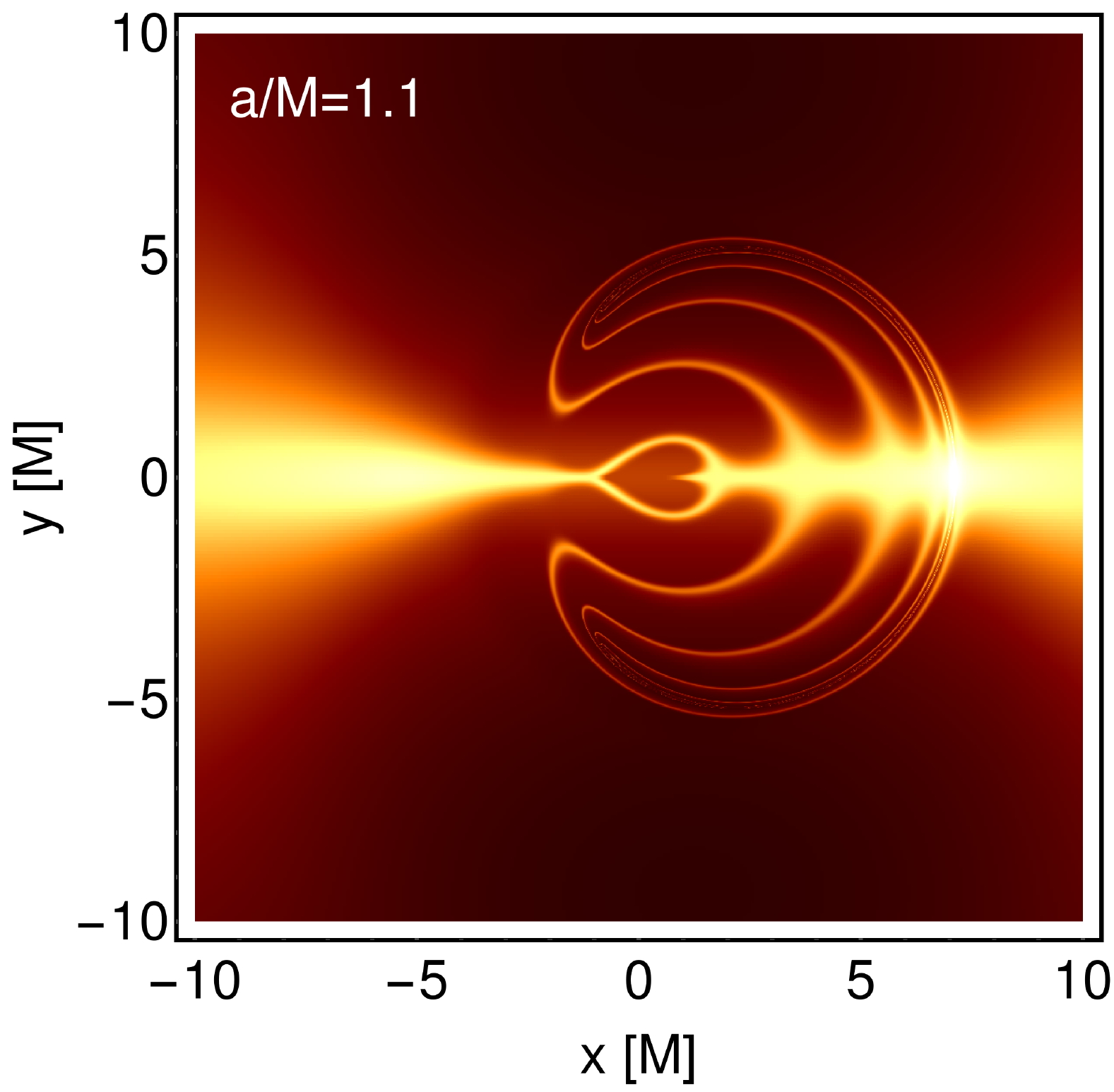}
	\hfill
	\includegraphics[width=0.325\linewidth]{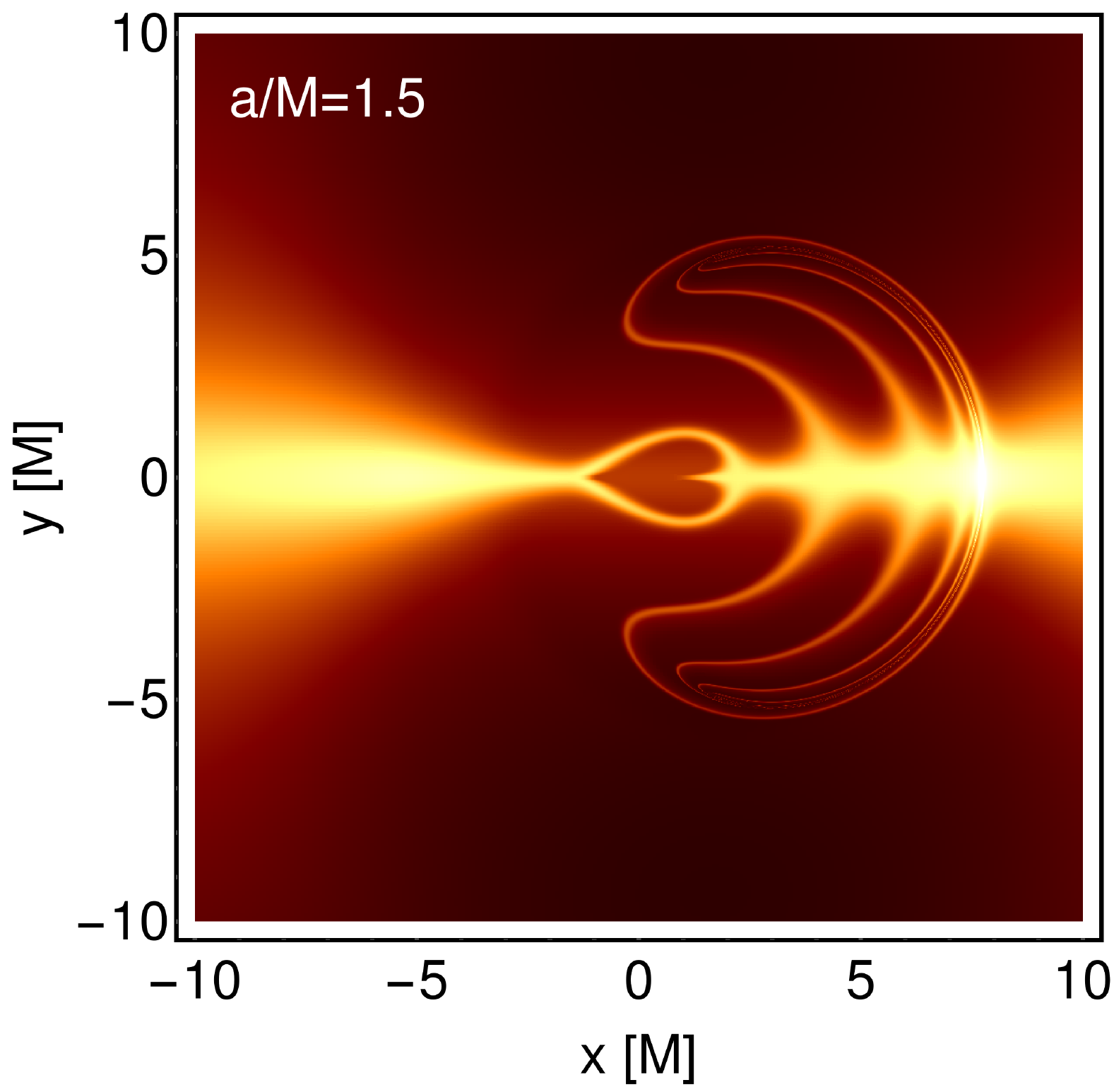}
	\hfill
	\includegraphics[width=0.325\linewidth]{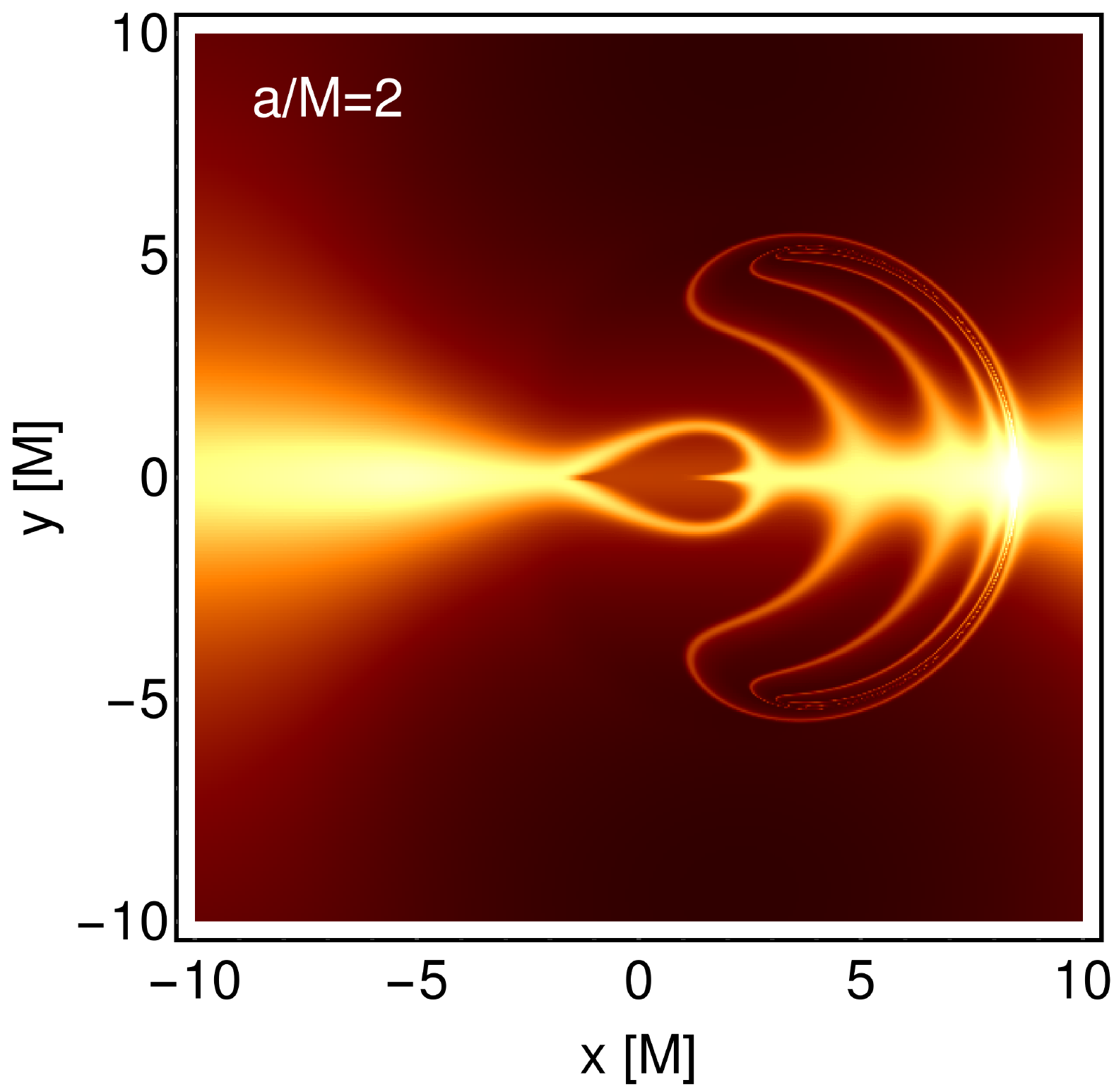}
\caption{\label{fig:superextremalPi2series}
As in Fig.~\ref{fig:superextremalM87series} but for edge-on observer inclination $\theta_\text{obs}=\pi/2$.
}
\end{figure}

We first explore how the simulated image changes as we increase the spin. As a baseline, we also show the image of a black hole, i.e., with horizon, at $a=0.99\,M< a_{\rm crit}$, cf.~upper left panels in Figs.~\ref{fig:superextremalM87series}, \ref{fig:superextremal8Pi20series} and \ref{fig:superextremalPi2series}. We consider three different inclinations, i.e., angles between the line of sight to the observer and the spin axis, namely close to face-on, $\theta_{\rm obs} = 17 \pi/180$ (inspired by the inclination of M87$^\ast$ \cite{CraigWalker:2018vam, paper1}) in Fig.~\ref{fig:superextremalM87series}, edge-on inclination ($\theta_{\rm obs} = \pi/2$) in Fig.~\ref{fig:superextremalPi2series} and an intermediate inclination of $\theta = 2 \pi/5$ in Fig.~\ref{fig:superextremal8Pi20series}.
These images show that the disk has relatively large extent in the ``slow model'' (cf.~Tab.~\ref{tab:parameters}), which is due to our choice of $\alpha$ and $r_{\rm cut}$. The $n=1$ ring is clearly visible for inclinations away from edge-on. In addition, higher-order photon rings are visible (with significantly lower intensity) and demarcate the shadow boundary. The characteristic asymmetry of the shadow due to nonzero spin is present away from near-face-on inclinations.

Starting from these baseline images, we increase the value of the spin parameter $a$. We consider values not far above the critical spin value, e.g., $a/M= 1+10^{-6}$. The value $a=M$, which is also already larger than $a_{\rm crit}$ is very challenging to access numerically.
Further, we increase the spin to significantly larger values all the way to $a=2\,M$.
These large values of $a>M$ may not necessarily be of phenomenological interest, but they are of theoretical interest, because of the intricate evolution of lensing features with $a$.\\

\begin{figure}[!t]
\centering
	\includegraphics[height=0.325\linewidth]{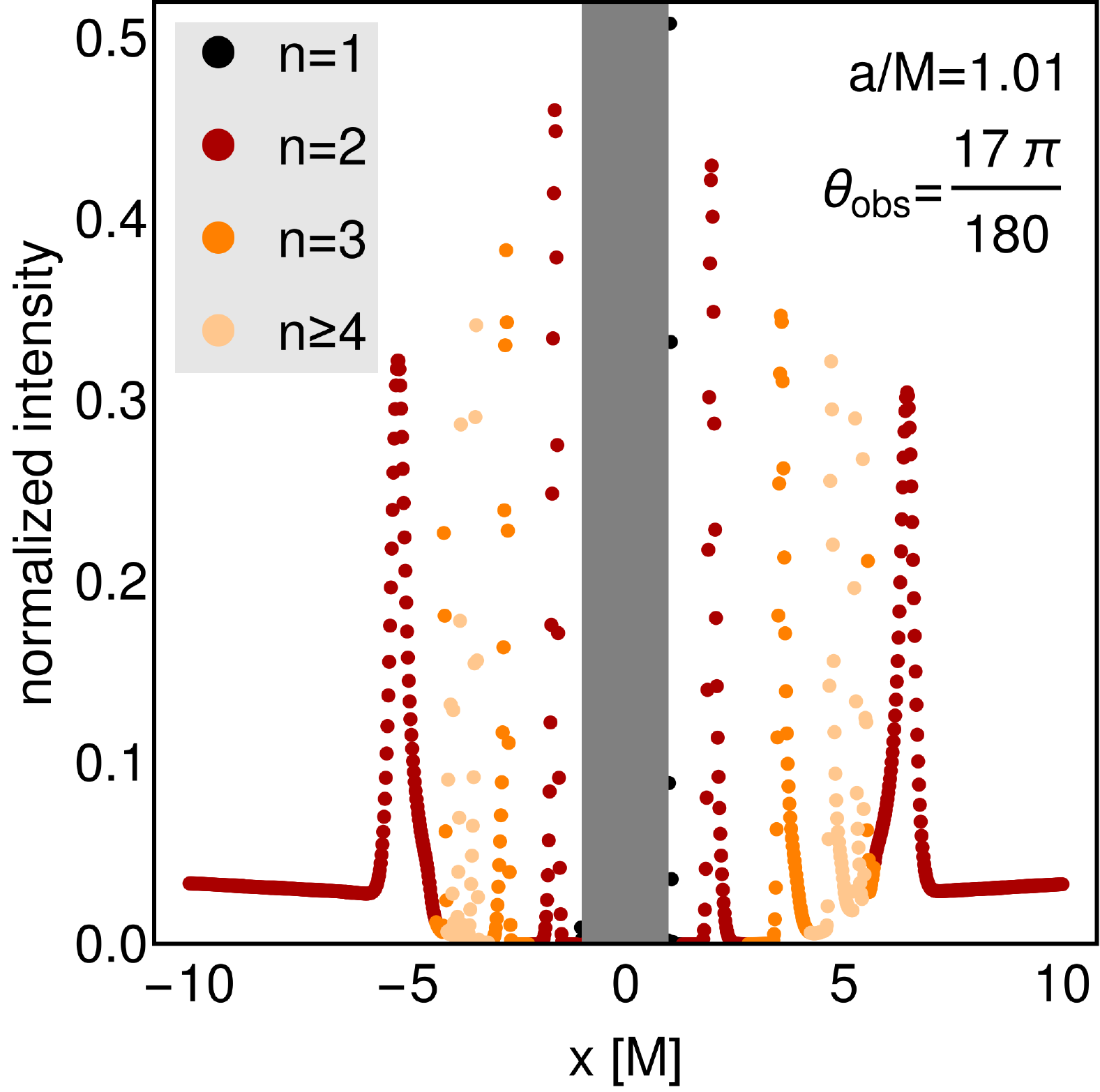}
	\hfill\vline\hfill
	\includegraphics[height=0.325\linewidth]{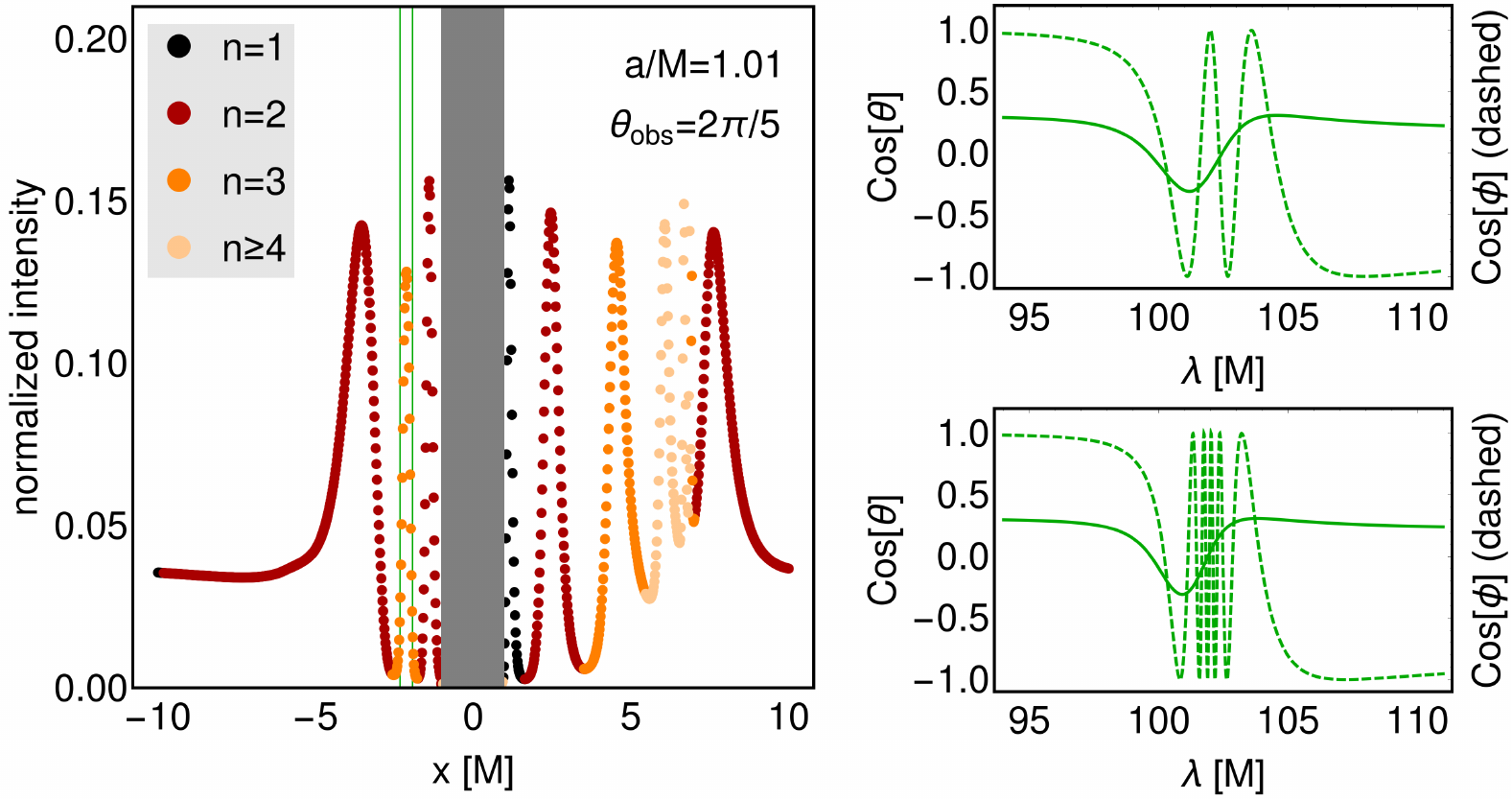}
\caption{\label{fig:crosssections}
We show normalized intensity cross-sections at fixed $y=0$ coordinate for varying $x$ coordinate in the image plane.
The spin-parameter is set to $a=1.01\,M$. In the left-hand (right-hand) panel the inclination of the observer's screen is set to $\theta_\text{obs}=17\pi/180$ ($\theta_\text{obs}=2\pi/5$). The intensity is normalized to the integrated intensity across the presented cross-section region. The different coloured points indicate the number of times $n_p$ that the trajectory pierces the $(\chi=0)$--plane (in which the disk is centered). In the right-hand panel, we also show (the cosine of) the coordinates $\theta$ and $\phi$ for two example trajectories (the image location of which is marked with green vertical lines in the main right-hand panel). While a unique piercing number $n_p$ can be assigned to each light ring, the winding number (i.e., oscillations in $\phi$) can vary within a single light ring, see also Fig.~\ref{fig:ringCounting}}.
\end{figure}

The lensing features that distinguish the black hole from the horizonless spacetime are due to trajectories that cross the would-be-location of the photon sphere. 
If a horizon is present, such trajectories do not reach asymptotic infinity. This disconnect of screenpoints to asymptotic infinity is the crucial global property that changes instantaneously (see Sec.~\ref{sec:lights-up}) once the horizon disappears.\\
As soon as the horizon disappears, trajectories incident on the screen region enclosed by the shadow boundary (critical curve)
of extremal Kerr spacetime can traverse the spacetime region within the photon sphere (of extremal Kerr spacetime). Hence, without horizon, they can have originated from asymptotic infinity.\\
As a result, new lensed image features appear in the image region enclosed by the shadow boundary (of extremal Kerr spacetime). These are labelled by the piercing number $n_p$. We have verified this numerically by counting the number of times that the respective trajectories traverse the equatorial plane, cf.~Fig.~\ref{fig:crosssections}. 
However, the horizonless case is different from the black-hole case in two important ways. 
First, each `external' photon ring labelled by $n_\text{ext}$ is accompanied by a twin `internal' photon ring labelled by $n_\text{int}$ with the same piercing number, i.e., $n\equiv n_\text{ext}\equiv n_\text{int}$. 

Second, for any $a>a_\text{crit}$, there is only a finite number of photon rings. (More precisely, this is the case for values of $a$ at which the photonsphere has disappeared, see Sec.~\ref{sec:lights-up}, which coincides approximately with $a_{\rm crit}$.)
The number of photon rings (both internal and external) drops with increasing $a$, starting from an infinite number of external photon rings at $a\leq a_{\rm crit}$.\\
With increasing superextremal spin (and for any non-face-on inclination), the twin pairs of photon rings of successively lower $n_p$ pinch together on the prograde image side. Once they pinch, they no longer form two distinct photon rings $n_\text{ext}$ and $n_\text{int}$ but rather combine into one highly lensed image of the accretion disk that traces the outline of a crescent.

There are two prerequisites for this annihilation of the twin-pairs of photon rings: First,
for inclinations away from face-on-inclination, the internal photon rings are centered around a point that is closer to the prograde than to the retrograde image side. Thus, they are closer to the external photon rings on the prograde image side. Second, the ordering of internal photon rings is reversed, compared to the ordering of external photon rings: The internal $n_p=1$ is the innermost ring, and internal rings get closer to the external rings, as $n_p$ increases. This can be seen in the images in Figs.~\ref{fig:superextremalM87series}, \ref{fig:superextremal8Pi20series} and \ref{fig:superextremalPi2series}, where the innermost internal ring is brighter and broader than inner rings that lie further outwards. Further, it can be seen in Fig.~\ref{fig:crosssections}, where we explicitly show the location of the internal rings and their labelling by $n_p$.
\\
Together, the fact that the internal ring with the largest $n_p$ lies very close to its external twin, and that they are closest on the prograde image side, enables the pinching mechanism to occur.

\subsection{Instantaneous lighting up}
\label{sec:lights-up}

\begin{figure}[!t]
\centering
	\includegraphics[width=0.49\linewidth]{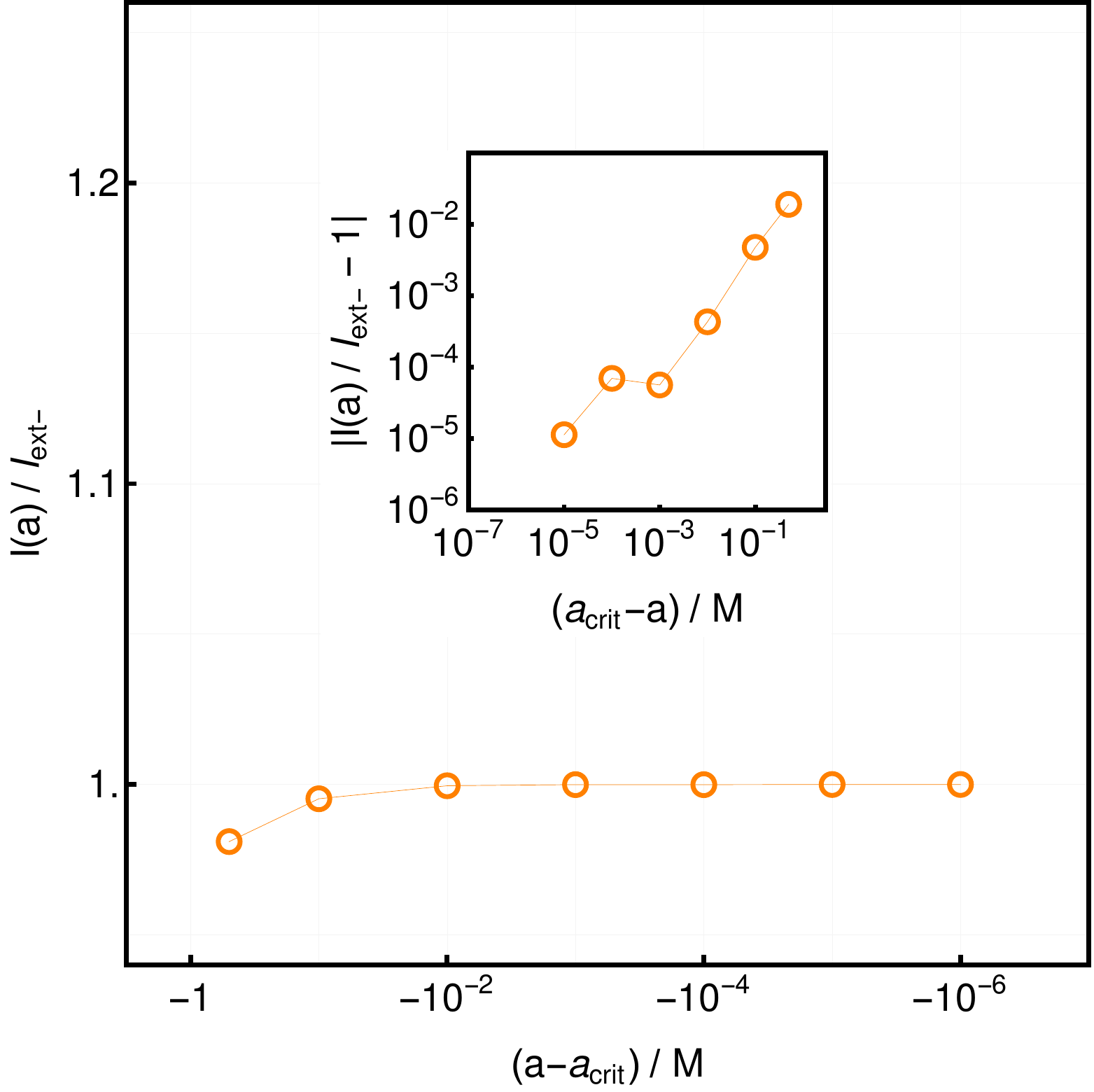}
	\includegraphics[width=0.49\linewidth]{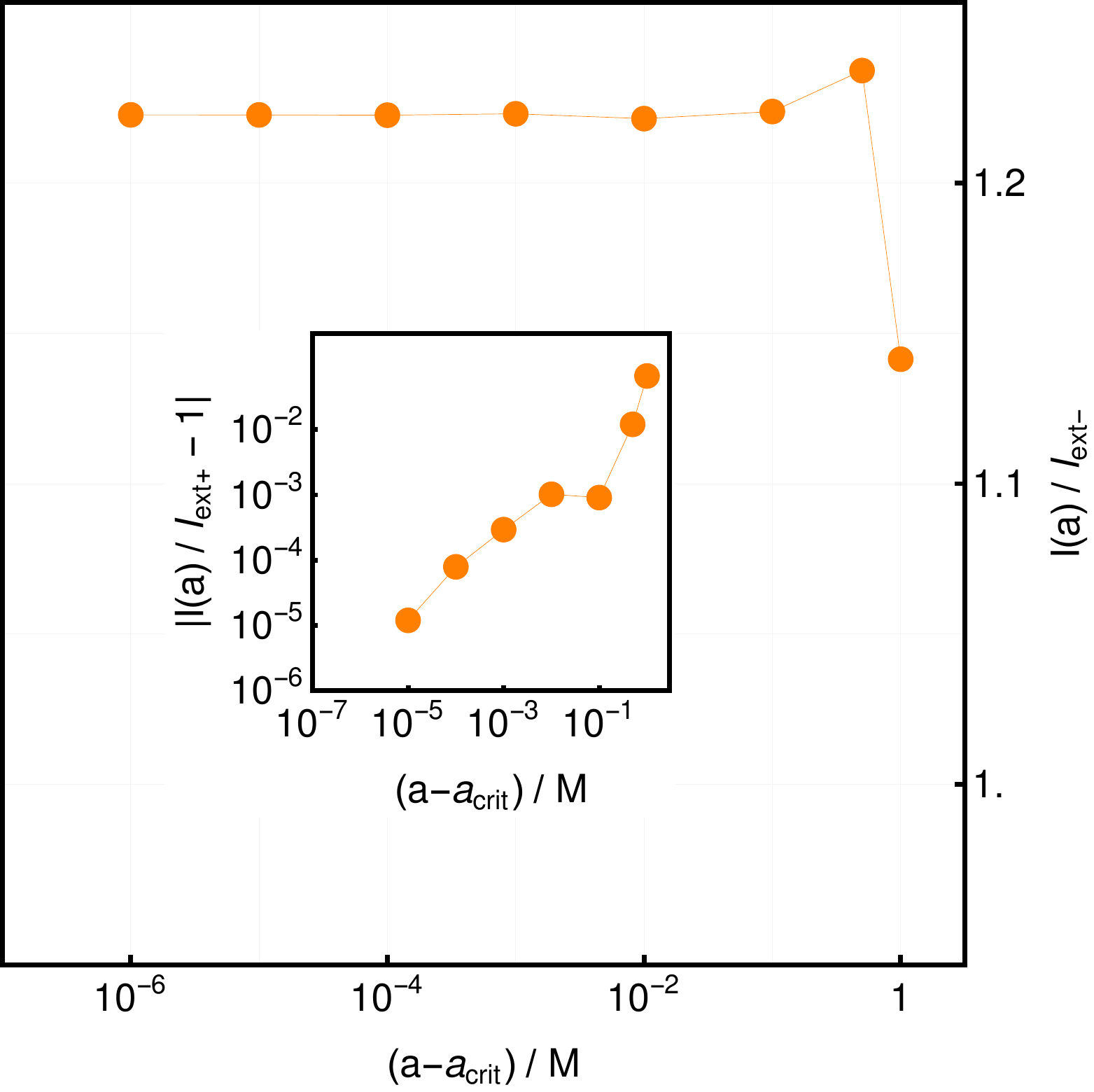}
\caption{\label{fig:lightsUp}
Spin parameter $a$-dependence of the image intensity $I(a)$ (integrated over the full image region $x,\,y\in~[-10M,\,10M]$) for inclination $\theta_\text{obs}=17\pi/180$, exhibiting discontinuous behaviour as the spin parameter $a$ approaches the critical value $a_\text{crit}$ from below (open circles in the left-hand panel) and from above (closed circles in the right-hand panel). 
\newline
In the main panels (both left and right), we normalize the intensity $I(a)$ to the same value, i.e., to the subextremal limit $I_{\text{ext}-} \approx \lim_{a\nearrow a_\text{crit}}I(a)$: This shows that the subextremal and superextremal limits disagree, i.e., that $\lim_{a\nearrow a_\text{crit}}I(a)\neq \lim_{a\searrow a_\text{crit}}I(a)$. 
In the left-hand and right-hand inset, we normalize to the respective subextremal and superextremal limit, i.e., to $I_{\text{ext}-}$ and $I_{\text{ext}+} \approx \lim_{a\searrow a_\text{crit}}I(a)$, respectively. This demonstrates the onset of power-law scaling.
In practice, we normalize to the value closest to criticality, i.e., to $I_{\text{ext}\pm} \equiv I(a=a_\text{crit}\pm10^{-7}M)$.
}
\end{figure}

We now explore the transition between a horizonless spacetime and a black hole and test whether the transition results in an instantaneous or a gradual change in the image. One may think of this in the language of phase transitions: defining an appropriate measure of image intensity in the central region as an order parameter, we ask whether the order parameter exhibits a continuous change (second-order transition) or a jump (first-order transition) when $a$ crosses $a_{\rm crit}$.
Given that the spacetime geometry (characterized, e.g., by invariants), changes only by $\mathcal{O}(\epsilon)$, when the spin is increased from $a$ to $a+ \epsilon$, one might expect that the transition from $a<a_{\rm crit}$ to $a>a_{\rm crit}$ is gradual. However, arguing in terms of the local geometry is misleading, because the disappearance of an event horizon is a global, not a local, change. At $a = a_{\rm crit} + \epsilon$, with arbitrarily small $\epsilon$, the event horizon is no longer present. Thus, null geodesics can travel through the spacetime region formerly inside the event horizon. These null geodesics add to the total image intensity and the additional lensed features besides the photon rings appear instantaneously at significant intensities. We support this claim by a numerical study, in which we determine the intensity $I(a)$, integrated across the full image region,
as a function of $a$. We define $I_{\rm ext-}$ as the intensity at the largest subcritical value in our set of simulations, i.e., at $a/M=1-10^{-6}$. This value approximates the limit $I(a)\overset{a \nearrow a_{\rm crit}}{\longrightarrow}I_{\rm ext-}$, i.e., it is the intensity associated to the black-hole case. Then, we
investigate $I(a)$ in relation to $I_{\rm ext -}$. If the transition between an image of a horizonless spacetime and a black hole was gradual, then the image intensity at supercritical spin would approach the intensity of the black-hole case in the limit where $a$ approaches $a_{\rm crit}$ from above, i.e., $I(a)/I_{\rm ext -} \overset{a \searrow a_{\rm crit}}{\longrightarrow}1$. In contrast, if the transition was instantaneous, then $I(a)/I_{\rm ext -} \overset{a \searrow a_{\rm crit}}{\longrightarrow}>1$, i.e., even at an epsilon above $a_{\rm crit}$, the total image intensity would be larger due to the presence of additional lensed features in the image. \\
First, we  confirm that $I(a)/I_{\rm ext -} \overset{a \nearrow a_{\rm crit}}{\longrightarrow} 1$ down to $a_{\rm crit}- a = 10^{-6}$. Because we observe an approximate power-law scaling with $(a_{\rm crit} - a)/M$, we consider the ratio to be apparently converged. Second, we calculate $I(a)/I_{\rm ext -} \overset{a \searrow a_{\rm crit}}{\longrightarrow}$ down to $a- a_{\rm crit} = 10^{-6}$. We again observe that the ratio is apparently converged.

Our main result in this section is that in the limit in which $a$ approaches $a_{\rm crit}$ from above, the ratio $I(a)/I_{\rm ext -}$ exceeds 1 significantly (by about 20 \%).
We expect that the numerical value varies with the disk model. The discrete jump may become more or less pronounced but remains present for any generic disk model. Thus,  for any infinitesimally supercritical $|a|>a_\text{crit}$, the image intensity and image features are markedly different from the image at subcritical spin. Therefore, the  effect that the black hole ``lights up'' is instantaneous.\\
In a thought experiment, in which a black hole is initially spinning at slightly subcritical spin and accretes matter with angular momentum, so that its spin exceeds $a_{\rm crit}$, our stationary model suggests that an instantaneous effect is visible in the image. Similarly, in an unphysical thought experiment, in which quantum effects could be switched on and off, a Kerr black hole spinning at $a_{\rm crit}<|a|<M$ would ``light up'' instantaneously, if quantum gravity was switched on. In that sense, quantum gravity would ``light up'' a black hole.\footnote{In a different sense, quantum gravity ``lights up'' black holes, because the transition to a horizonless spacetime is forbidden in GR by cosmic censorship, but not forbidden by a generalized cosmic censorship conjecture which forbids the existence of naked singularities also beyond GR. Thus, the presence of singularity-resolving quantum-gravity effects makes the ``lighting up'' process possible.}

\section{What would the (ng)EHT see?}
\label{sec:(ng)EHT}

\begin{figure}[!t]
\centering
	\includegraphics[width=\linewidth]{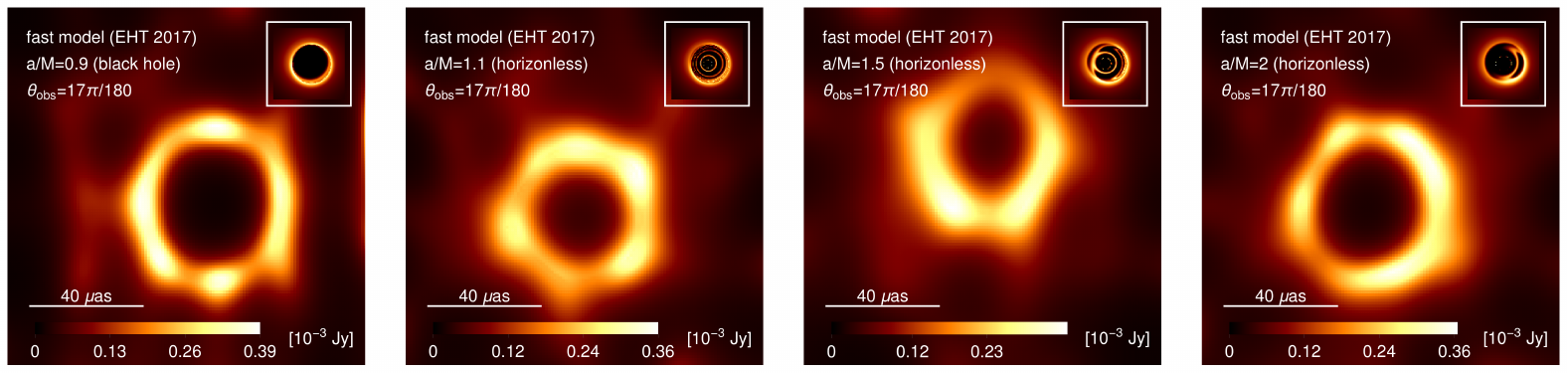}
	\includegraphics[width=\linewidth]{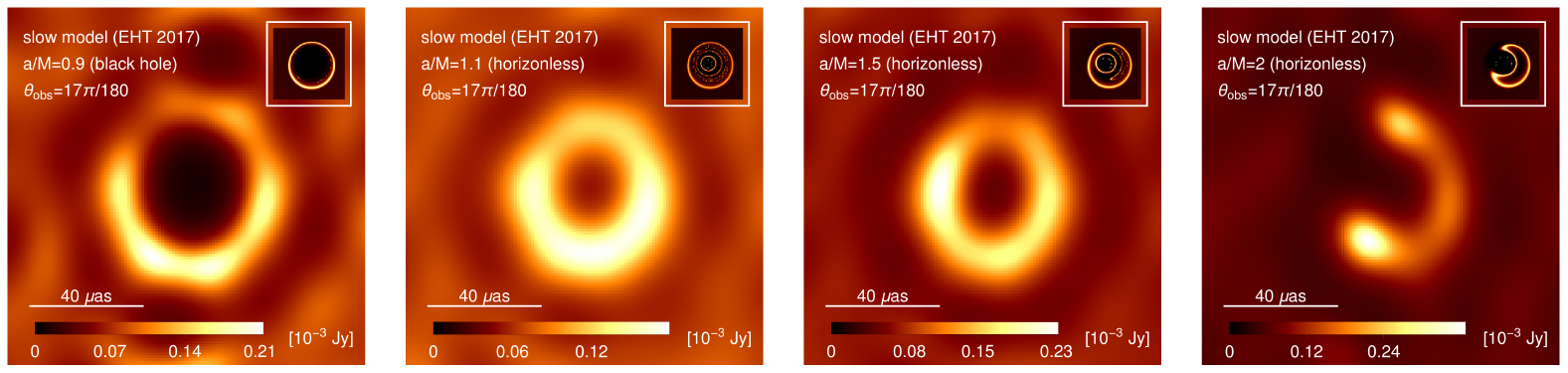}
	\caption{\label{fig:increasingSpin_EHT2017}
	We show the ehtim~\cite{Chael:2018oym} image reconstructions with the EHT 2017 array at M87* inclination: From left to right, we show the black-hole case ($a=0.9\,M$) and increasingly overspun horizonless spacetimes ($a=1.1\,M$, $a=1.5\,M$, and $a=2\,M$). The colorbar indicates the flux (in Jy) at each pixel. For reference, we show (a slightly zoomed in version of) the high-resolution image in the top-right corner of each panel, see also Fig.~\ref{fig:superextremalM87series}. In the upper (lower) row, we show reconstructions of simulated images using the disk model with fast (slow) asymptotic falloff, cf.~Tab.~\ref{tab:parameters}. (The $(x,y)$-displacement between the reconstructed images is an artefact of the reconstruction and has no physical meaning.)
	}
\end{figure}
\begin{figure}[!t]
\centering
	\includegraphics[width=\linewidth]{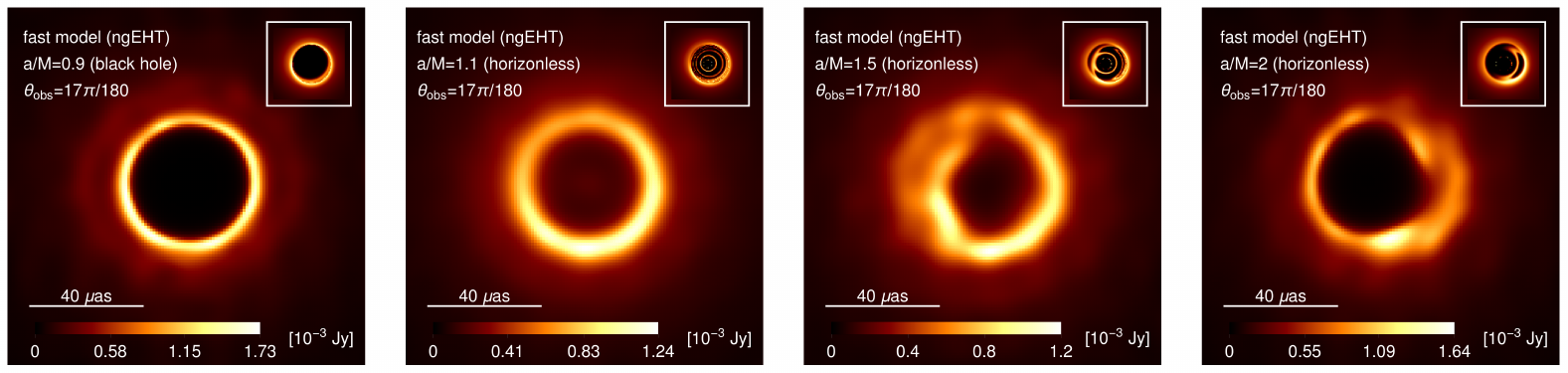}
	\includegraphics[width=\linewidth]{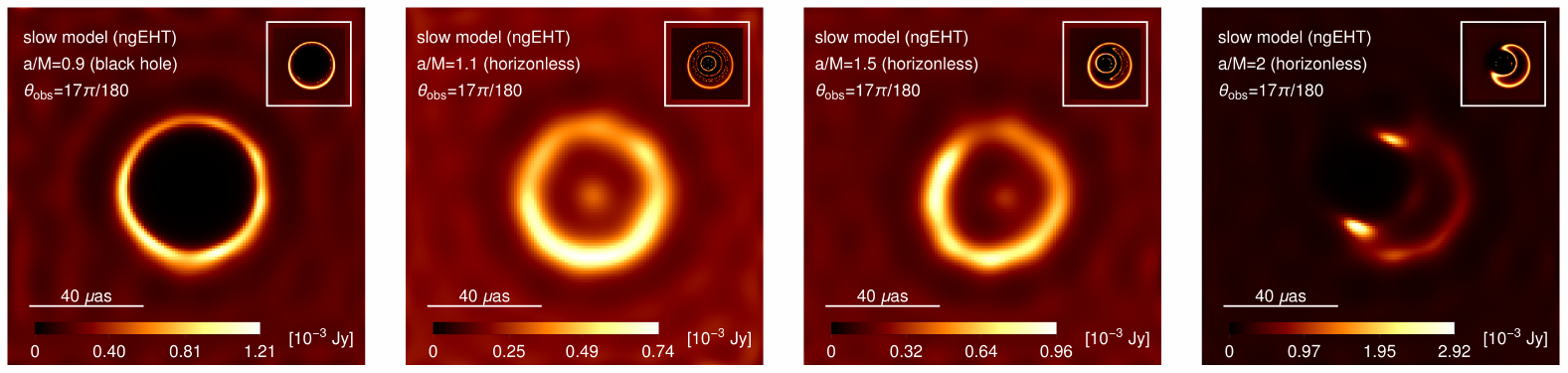}
	\caption{\label{fig:increasingSpin_futureArrays}
	As in Fig.~\ref{fig:increasingSpin_EHT2017} but for a potential ngEHT multifrequency (at 230 and 345 GHz) observation with 20 telescopes, see~\cite[App.~A]{Eichhorn:2022fcl} for details of the telescope configuration.
	}
\end{figure}

The simulated images of horizonless spacetimes look rather distinct from black-hole images, at least at the high resolution shown in Sec.~\ref{sec:lights-up}. However, the EHT is a sparse array, with the existing baselines covering only parts of the Fourier plane. In particular, the resolution is limited by the diffraction limit $\sim\lambda/D$, where $\lambda$ is the observing wavelength and $D$ the baseline. Therefore, reconstructed images have a lower resolution than those shown in Sec.~\ref{sec:lights-up} (at least if the object in question has mass parameter and distance roughly as M87*). Hence, EHT measurements might not be sensitive and resolved enough to detect all characteristic image features of a horizonless spacetime. Indeed, in \cite{Eichhorn:2022fcl} it has been shown that reconstructed images of overspun spacetimes with $a<1.01$, similarly to images of wormholes \cite{Lamy:2018zvj,Vincent:2020dij}, mainly differ from images of Kerr black holes at the 2017 EHT resolution, because the central brightness depression is less pronounced. 
For $a > 1.01$, image structures are different and thus it is worthwhile to investigate whether the same conclusion holds. This question is of interest even if the spin values we consider in this section may not be achievable in astrophysical settings, because the simulated images provide useful templates of distinct image structures in which the capabilities of VLBI arrays can be tested.

To investigate this question, we use the ehtim toolkit~\cite{Chael:2018oym}. 
We work with images obtained with both the ``slow model'' and the ``fast model'', cf.~Tab.~\ref{tab:parameters}.
The resulting high-resolution simulated images serve as input in the ehtim toolkit~\cite{Chael:2018oym} which  simulates an  observation by a specified telescope configuration. We reconstruct images for (i) the EHT 2017 array with 7 active telescopes and a single observing frequency of $230$ GHz and (ii) for a potential future ngEHT array with 20 active telescopes observing simultaneously at $230$ and $345$ GHz. Information from both frequencies is combined in the reconstruction and we assume a spectral index of $\alpha=1.88$~\cite{EventHorizonTelescope:2021dvx}.
The telescope specifications can be found in \cite[App.~A]{Eichhorn:2022fcl}, see also~\cite{ngEHTexplorer.git}. All fits, generated by the ehtim toolkit and presented in this paper, come with a $\chi^2$-value in the range $0.89<\chi^2<1.08$.
\\

First, we focus on the observationally relevant case of near face-on inclination of M87*, i.e., $\theta_\text{obs}=17\pi/180$. The latter is known due to independent observations of the jet~\cite{paper5}. The EHT observation of Sgr A* is also more compatible with simulated images at a (near) face-on inclination~\cite{EventHorizonTelescope:2022urf}. 

For the EHT 2017 array, reconstructed images (see Fig.~\ref{fig:increasingSpin_EHT2017}) of the black-hole case (at spin $a=0.9\,M$) and the horizonless cases appear similar because the inner photon rings, which appear once the horizon is absent, remain unresolved. Even for the significantly overspun cases up to $a=1.5 M$ no significant difference is visible by eye. As in \cite{Eichhorn:2022fcl}, the central brightness depression may differ between the black-hole case and the horizonless cases, however, these differences are likely not highly significant at the limited dynamic range of the 2017 EHT array. 

In the image at $a=2$, the conclusions drawn from the ``slow model'' and the ``fast model'' differ, because the simulated images differ significantly: both models feature the $n=1$ photon ring in the form of a crescent. For the ``slow model'', the remaining $n=0$ emission is distributed throughout the field of view, whereas most of the $n=0$ emission is concentrated in a ring at the inner disk cutoff for the ``fast model''. This difference between a crescent and a crescent embedded in a ring is resolved by the EHT 2017 configuration, as one would expect based on the effective resolution of that array.

In \cite{Eichhorn:2022fcl} it has been shown, that images at $a=1.01$ with the ``fast model'' are similar to the 2017 EHT data on M87* on large scales in the Fourier plane, while there is a deviation at small scales in the Fourier plane. These small scales in the Fourier plane correspond to large distances in the image domain. At these distances, comparing the reconstructed images by eye, cf.~Fig.~\ref{fig:increasingSpin_EHT2017}, we see that the ``fast model'' performs better, because the ``slow model'' features significant image intensity throughout the EHT field of view, unlike the actual image of M87*. Nevertheless, significant differences of the reconstructed image (apart from differences in the diffuse emission) only appear at $a=2$, where the image topology of the simulated image differs between the ``slow model'' and the ``fast model''.

The situation changes with the capabilities of a potential ngEHT configuration, at least for the specific telescope configuration investigated here (see Fig.~\ref{fig:increasingSpin_futureArrays}): in all of the significantly overspun cases, i.e., $a\gtrsim1.5\,M$, the reconstructed images deviate visibly from the black-hole case $a=0.9 M$. The crescent-like structures that appear in the simulated images due to the merging of the inner and outer photon rings lead to an asymmetry of the ring in the reconstructed images. For the ``slow model'', differences in the central brightness depression appear even earlier. This is probably because more of the diffuse emission is distributed throughout the field of view; thus, the intensity difference between the ring and the center is lower and, at a given dynamic range,  central image structures can more easily be reconstructed.
The case of marginally overspun black holes, i.e., $a<1.1$ is more subtle (compare the two left panels in Fig.~\ref{fig:increasingSpin_futureArrays}) and is investigated separately in~\cite{Eichhorn:2022fcl}.
\\
\begin{figure}[!t]
\centering
	\includegraphics[width=\linewidth]{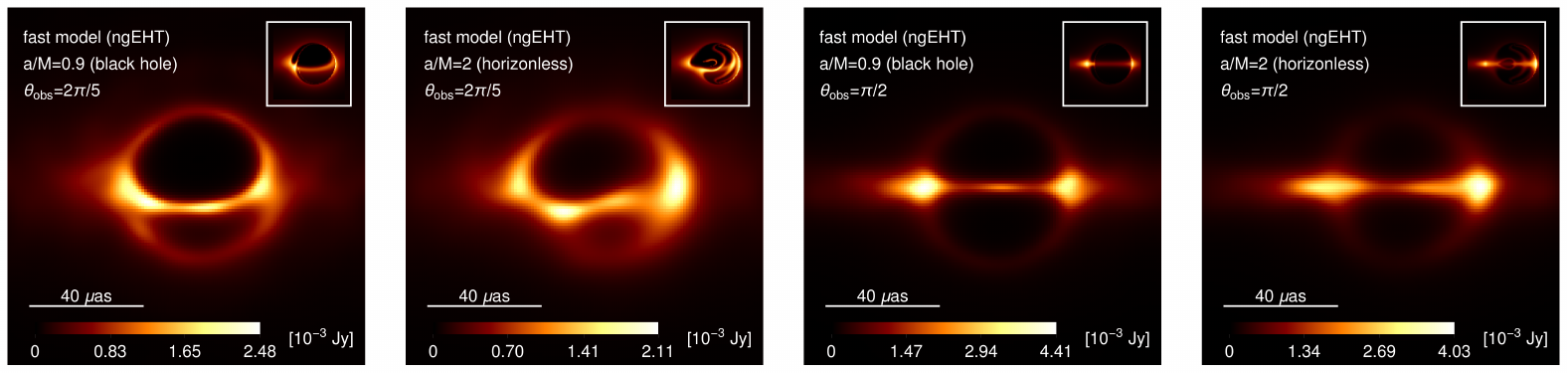}
	\includegraphics[width=\linewidth]{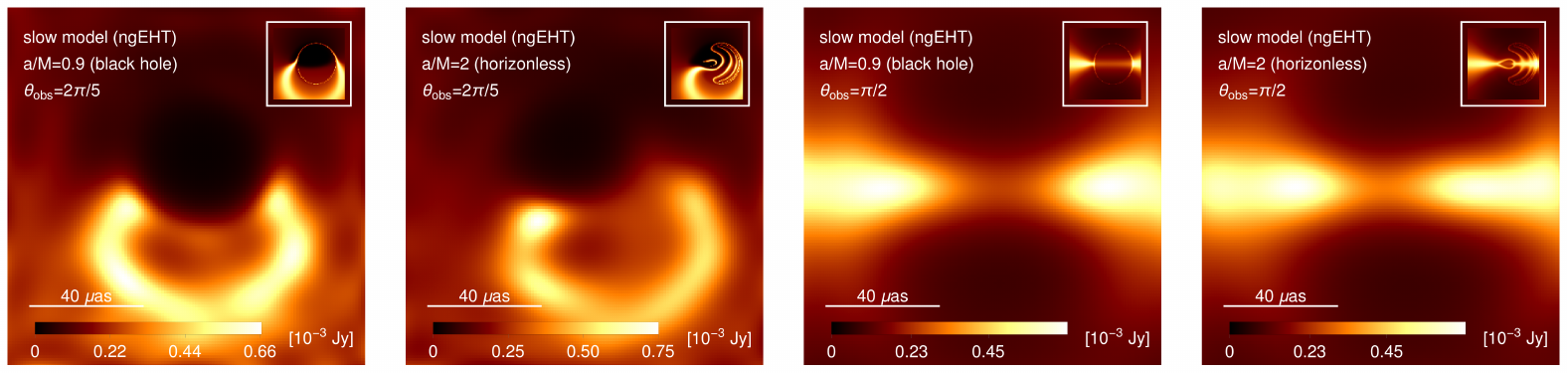}
	\caption{\label{fig:inclinationSeries_futureArrays}
	We show the ehtim~\cite{Chael:2018oym} image reconstructions for a potential ngEHT multifrequency (at 230 and 345 GHz) observation with 20 telescopes, see~\cite[App.~A]{Eichhorn:2022fcl} for details: In the two left panels (two right panels), we compare the black-hole case ($a=0.9\,M$) with an overspun horizonless spacetimes ($a=2\,M$), for near edge-on ($\theta_\text{obs}=2\pi/5$) and edge-on ($\theta_\text{obs}=\pi/2$) inclination, respectively.
	In the upper (lower) row, we show reconstructions of images using the disk model with fast (slow) asymptotic falloff, cf.~Tab.~\ref{tab:parameters}.
	The colorbar indicates the flux (in Jy) at each pixel. For reference, we show the high-resolution image in the top-right corner of each panel, see also Fig.~\ref{fig:superextremal8Pi20series} and~\ref{fig:superextremalPi2series}.
	}
\end{figure}

We also show reconstructions at (near) edge-on inclination, see Fig.~\ref{fig:inclinationSeries_futureArrays}. These are less relevant for a comparison to the 2017 EHT observations of M87* and Sgr A*, but serve as a useful theoretical example. It turns out that in those cases it is even harder to distinguish the black hole from the horizonless case. In Fig.~\ref{fig:inclinationSeries_futureArrays}, we compare the reconstructed ngEHT images of black holes and horizonless spacetimes for $\theta_\text{obs}=2\pi/5$ and $\theta_\text{obs}=\pi/2$. In contrast to simulated observations at (near) face-on inclination, the respective images at (near) edge-on inclination remain qualitatively similar for both disk models. Thus it is fortunate that M87* and Sgr A* have (near) face-on inclination because future observations may constrain horizonless spacetimes more easily.

\section{Conclusions and outlook}\label{sec:conclusions}

We discuss a scenario in which quantum-gravity effects on astrophysical black holes may be observable, even if quantum gravity is strictly tied to Planckian curvature scales.
The spin parameter of a regular spacetime serves as a lever arm that translates Planck-suppressed (and thus tiny) local modifications of the spacetime into significant
changes in the global structure of the spacetime, namely the presence or absence of a horizon. Spacetimes with and without horizon can look qualitatively distinct to a radio VLBI observer (given appropriate resolution and dynamic range):
whereas they cannot see any light that would have to traverse the horizon on its way from the accretion disk to their detector, that same light ray can make it to their detector, when the horizon is absent. 

We work with a particular regular spinning spacetime, based on a locality principle~\cite{Eichhorn:2021etc,Eichhorn:2021iwq}. This spacetime can also be motivated from the physics of asymptotically safe quantum gravity~\cite{Bonanno:1998ye,Bonanno:2000ep,Bonanno:2006eu,Falls:2010he,Falls:2012nd,Litim:2013gga,Koch:2013owa,Koch:2014cqa,Kofinas:2015sna,Torres:2017gix,Torres:2017ygl,Pawlowski:2018swz,Adeifeoba:2018ydh,Platania:2019kyx,Held:2019xde}. 
Just like for a Kerr black hole, the spin parameter $a$ determines the presence or absence of an event horizon. Above a critical value of the spin parameter $a_{\rm crit}< M$, the horizon disappears. One may thus imagine two transitions from a black hole to a horizonless spacetim: (i) a purely theoretical transition: At $|a|=a_{\rm crit}$, the quantum-gravity effects are switched on, starting from
a Kerr black hole, leading to a loss of the horizon; (ii) a potentially astrophysical transition: In the presence of quantum-gravity effects, overspinning the black hole to $|a|>a_{\rm crit}$ also leads to a loss of the horizon.
In this transition, the shadow cast by the spacetime changes and additional light reaches an observer, thus the black hole ``lights up''.
\\

The value of the critical spin parameter $a_\text{crit}$ at which the black hole `lights up' depends on specifics of the quantum gravity effects, but lies below $a_{\rm crit,\, Kerr}= M$, as long as quantum gravity counteracts geodesic focusing. For the class of regular spinning spacetimes 
we consider, its value is determined by the behavior of an effective mass function.
For the regular spinning spacetime inspired by asymptotically safe quantum gravity, we find $a_{\rm crit} \approx M \left(1- 4\sqrt{3}\,\ell_{\rm Planck}^2/M^2 \right)+\mathcal{O}(\ell_{\rm Planck}^4/ M^4)$. Hence, there exist classical black-hole spacetimes with near-extremal spin which correspond to horizonless spacetimes once quantum-gravity effects are taken into account. Whether or not astrophysical black holes can `light up' thus depends on whether or not there exists a physical process that can drive black holes to these near-extremal spin parameters, cf.~Sec.~\ref{sec:superspinning-review}. We stress that this is a distinct question from overspinning a Kerr black hole, which requires $|a|>M$ and can be explored in GR. In contrast, we require  $a$ such that $a_{\rm crit}<|a|<M$ and work in a setting beyond GR.
\\

We investigate how the image of such a spacetime changes across the transition from a black hole to a horizonless spacetime, by (i) characterizing the new image features of the horizonless spacetimes, (ii) investigating whether the black holed ``lights up'' gradually or instantly and (iii) simulating observations with EHT and ngEHT arrays to test whether the transition may actually be detectable.
To do so, we use an analytical disk model of an optically and geometrically thin accretion disk. We simulate the resulting shadow images by numerical integration of the radiative transfer equation along ray-traced geodesics incident on an observer's screen at asymptotic infinity.

We find that the images of horizonless spacetimes show a finite set of photon rings. Photon rings arise not as a consequence of a horizon, but merely due to a photon sphere, i.e., a closed surface of marginally bound photon orbits. Even an approximate photon sphere (in which the photon-sphere equation has a formal solution with a small imaginary part) can still result in a successive (although not infinite) set of photon rings.
In the presence of a horizon, there is only one set of photon rings which corresponds to geodesics which approximate the photon sphere from the outside. Any geodesic which approximates the photon sphere from the inside is intercepted by the horizon. This changes in the absence of a horizon: A second set of `inner' photon rings is formed by geodesics which approximate the photon sphere from within. We demonstrate that both the `outer' and `inner' photon rings can be labelled by a piercing number which counts the number of times the respective orbits pierce through the accretion disk, cf.~Fig.~\ref{fig:crosssections}. \\
As the spin parameter increases (and the imaginary part of the photon-ring location grows) the inner and outer photon rings annihilate pairwise by pinching off and forming crescents which subsequently shrink.
\\
Due to the appearance of the inner photon rings, the black hole ``lights up'' and the total image intensity increases. We find that this effect is instantaneous, by comparing the total image intensity in the two limits $a \nearrow a_{\rm crit}$ and $a \searrow a_{\rm crit}$ and finding a significant difference.
Since the deviation of $a_\text{crit}$ from $a_\text{crit, Kerr}\equiv M$ is tiny, i.e., $|a_\text{crit, Kerr} - a_\text{crit}| \sim \ell_\text{Planck}^2/M^2 \sim 10^{-96}$ for M87*, the numerics to obtain reliable geodesics becomes increasingly expensive as $a\rightarrow a_\text{crit}$. While we cannot numerically explore $a\equiv a_\text{crit}$ directly, we find that the total flux scales to appreciably distinct values on either side of the transition, cf.~Fig.~\ref{fig:lightsUp} and Sec.~\ref{sec:lights-up}. While the precise percentage of additional flux depends on the specific disk model, the fact that the black hole `lights up' instantaneously (i.e., for any marginally supercritical spin parameter $a>a_\text{crit}$), does not.

Finally, we take a qualitative step to explore the prospects of using VLBI observations to constrain the presence of `inner' photon rings and the respective excess interior brightness. To do so, we simulate observations with the EHT 2017 telescope array as well as a potential ngEHT array configuration, cf.~Sec.~\ref{sec:(ng)EHT}. We compare black-hole images (at $a=0.9 M$) with those of horizonless spacetimes of increasingly supercritical spin parameter ($a=1.1\,M$, $a=1.5\,M$, $a=2\,M$). When reconstructed with the EHT 2017 array images of the black hole and the horizonless case appear similar, because the interior structures cannot be resolved. The only exception to this is the case $a =2 M$ with one of the two disk models that we consider. This value of $a$ is clearly of a more academic than phenomenological interest.
We find that the improved capabilities of an ngEHT array may suffice to resolve some of the image structures at significantly supercritical spin. A more quantitative analysis of the ngEHT capabilities with respect to these cases therefore is a promising avenue for future work. In particular, calculating the central brightness depression in a template-based procedure, as in \cite{Eichhorn:2022fcl} and comparing simulated data in the Fourier domain are two ways in which the capabilities of various arrays can in the future be determined more quantitatively.
\\

Overall, we conclude that the process of ``lighting up'' a black hole through quantum-gravity effects -- if it can indeed occur in an astrophysical setting for a SMBH -- has the prospect of being detectable by future radio VLBI arrays. If a future array would indeed observe a change in the image, of, e.g., M87* that is compatible with a transition from a black hole to a horizonless spacetime, more studies would of course be needed to investigate whether a similar change in appearance cannot be achieved in any other way, e.g., through significant changes in the accretion flow. At the same time, such an observation would motivate a strong theoretical effort to go beyond the toy model we consider here and construct a spinning black hole/horizonless spacetime completely free from pathologies and in which the fate of geodesics and infalling matter in the (Planckian) core region is well understood. This calls for significant developments within quantum gravity, in which such questions cannot comprehensively be answered yet.
\\

\emph{Acknowledgements:} We thank R.~Gold for discussions.
A.~E.~is supported by a research grant (29405) from VILLUM fonden. 
The work leading to this publication was supported by the PRIME programme of the
German Academic Exchange Service (DAAD) with funds from the German Federal Ministry of Education and Research (BMBF).

\bibliography{References}

\end{document}